\documentclass[10pt,aps,physrev,twocolumn,amsfonts,floatfix,superscriptaddress,accepted=2023-03-20,a4paper]{quantumarticle}

\pdfoutput=1

\usepackage[utf8]{inputenc}
\usepackage{amsmath,amssymb,amsfonts,amsthm}
\usepackage{mathtools}
\usepackage{graphicx}
\usepackage[caption=false]{subfig}
\usepackage{enumerate}
\usepackage{braket}
\usepackage{tikz}
\usetikzlibrary{intersections,shapes.arrows,calc}
\usepackage{quantikz}
\usepackage{tikz-network}
\usepackage[english]{babel}
\usepackage[ruled,linesnumbered]{algorithm2e}
\usepackage{rotating}
\usepackage[pdftex,bookmarks=false,colorlinks=true,linkcolor=blue,citecolor=blue,filecolor=black,urlcolor=blue]{hyperref}
\usepackage{datetime}

\DeclarePairedDelimiter\abs{\lvert}{\rvert}

\graphicspath{
    {figures/}
}

\DeclareMathOperator{\e}{e}

\newcommand{\C}{\mathbb{C}}

\definecolor{mblue}  {rgb}{0.368417,0.506779,0.709798}
\definecolor{mred}   {rgb}{0.922526,0.385626,0.209179}
\definecolor{mgreen} {rgb}{0.560181,0.691569,0.194885}
\definecolor{morange}{rgb}{0.880722,0.611041,0.142051}
\definecolor{mpurple}{rgb}{0.647624,0.37816,0.614037}
\definecolor{mcyan}  {rgb}{0.363898,0.618501,0.782349}
\newcommand{\nodecolor}{mblue!33!white}
\newcommand{\gatecolor}{mred!75!white}

\begin{document}

\title{Simulating quantum circuits using tree tensor networks}

\author{Philipp Seitz}
\email{philipp.seitz@tum.de}
\affiliation{Technical University of Munich, Department of Informatics, Boltzmannstra{\ss}e 3, 85748 Garching, Germany}

\author{Ismael Medina}
\email{ismael.medina@cs.uni-goettingen.de}
\affiliation{University of G\"ottingen, Campus Institute Data Science}

\author{Esther Cruz}
\email{esther.cruz@mpq.mpg.de}
\affiliation{Max-Planck-Institute of Quantum Optics, Hans-Kopfermann-Stra{\ss}e 1, 85748 Garching, Germany}

\author{Qunsheng Huang}
\email{keefe.huang@tum.de}
\affiliation{Technical University of Munich, Department of Informatics, Boltzmannstra{\ss}e 3, 85748 Garching, Germany}

\author{Christian B.~Mendl}
\email{christian.mendl@tum.de}
\affiliation{Technical University of Munich, Department of Informatics, Boltzmannstra{\ss}e 3, 85748 Garching, Germany}

\newdate{date}{22}{03}{2023}
\date{\displaydate{date}}

\begin{abstract}
We develop and analyze a method for simulating quantum circuits on classical computers by representing quantum states as rooted tree tensor networks. Our algorithm first determines a suitable, fixed tree structure adapted to the expected entanglement generated by the quantum circuit. The gates are sequentially applied to the tree by absorbing single-qubit gates into leaf nodes, splitting two-qubit gates via singular value decomposition and threading the resulting virtual bond through the tree. We theoretically analyze the applicability of the method as well as its computational cost and memory requirements and identify advantageous scenarios in terms of required bond dimensions as compared to a matrix product state representation. The study is complemented by numerical experiments for different quantum circuit layouts up to $37$ qubits.
\end{abstract}

\maketitle

\section{Introduction}
Tensor networks provide a well-established framework and method for analyzing and simulating strongly correlated quantum systems \cite{Verstraete2008, Schollwoeck2011, Bridgeman2017}.
Recently, these methods have been adapted to circuit-based (digital) quantum computers, either by representing the statevector as a matrix product state (MPS) \cite{Zhou2020}, or interpreting the overall quantum circuit as a tensor network. Such networks are contracted with the aid of heuristics to obtain a good contraction order and ``Feynman-simulator''-type delayed contractions \cite{Pan2020, Huang2020, Peng2020, Gray2021, Pan2021, Lykov2020}.

In this work, we aim to combine the advantages of both approaches by representing statevectors as tree tensor networks (TTNs).
The properties and capabilities of such tree-type tensor networks have been studied in detail in the past.
In particular, the ``multi-scale entanglement renormalization ansatz'' (MERA) \cite{Vidal2007, Vidal2008} can efficiently describe critical ground states of one-dimensional systems.
The use of TTNs to simulate strongly correlated chemistry systems \cite{Murg2010, Gerster2014, Murg2015, Gunst2018, Schroeder2019}, two-dimensional quantum systems \cite{Tagliacozzo2009}, and tensor differential equations \cite{Ceruti2021} has been investigated, but a dedicated study and analysis in the domain of digital quantum computers are still missing.
A similar study on hybrid tensor networks (including classical and quantum tensors) is done in \cite{Yuan2021} which includes some aspects of hybrid TTNs. 
Ref.~\cite{Dumitrescu2017} studies the entanglement properties of Shor's algorithm on tree tensor networks specifically.
Many useful properties of TTNs have been studied in the context of many-body physics \cite{Gerster2014}.
Optimizing tree layouts based on the system characteristics is just one of these, which we aim to apply to circuit simulation. 

\begin{figure}
\centering
\begin{tikzpicture}[>=stealth, scale=0.9]
    \node[circle, draw, thick, minimum size=12, inner sep=0, fill=\nodecolor] (C) at (-3,   0) {};
    \node[circle, draw, thick, minimum size=12, inner sep=0, fill=\nodecolor] (D) at (-2,   0) {};
    \node[circle, draw, thick, minimum size=12, inner sep=0, fill=\nodecolor] (A) at (-1,   0) {};
    \node[circle, draw, thick, minimum size=12, inner sep=0, fill=\nodecolor] (E) at ( 0,   0) {};
    \node[circle, draw, thick, minimum size=12, inner sep=0, fill=\nodecolor] (F) at ( 1,   0) {};
    \node[circle, draw, thick, minimum size=12, inner sep=0, fill=\nodecolor] (B) at ( 2,   0) {};
    \node[circle, draw, thick, minimum size=12, inner sep=0, fill=\nodecolor] (G) at ( 3,   0) {};
    \node[circle, draw, thick, minimum size=12, inner sep=0, fill=\nodecolor] (O) at (-2.5, 1.25) {};
    \node[circle, draw, thick, minimum size=12, inner sep=0, fill=\nodecolor] (P) at (-0.5, 1.25) {};
    \node[circle, draw, thick, minimum size=12, inner sep=0, fill=\nodecolor] (Q) at ( 2,   1.25) {};
    \node[circle, draw, thick, minimum size=12, inner sep=0, fill=\nodecolor] (R) at ( 0,   2.5) {};
    \draw[thick] (-3,-0.6) -- (C);
    \draw[thick] (-2,-0.6) -- (D);
    \draw[thick] (-1,-0.6) -- (A);
    \draw[thick] ( 0,-0.6) -- (E);
    \draw[thick] ( 1,-0.6) -- (F);
    \draw[thick] ( 2,-0.6) -- (B);
    \draw[thick] ( 3,-0.6) -- (G);
    \draw[thick] (C) -- (O);
    \draw[thick] (D) -- (O);
    \draw[thick] (A) -- (P);
    \draw[thick] (E) -- (P);
    \draw[thick] (F) -- (Q);
    \draw[thick] (B) -- (Q);
    \draw[thick] (G) -- (Q);
    \draw[thick] (O) -- (R);
    \draw[thick] (P) -- (R);
    \draw[thick] (Q) -- (R);
    \draw[decorate, decoration={brace,amplitude=6}] ( 3.4, 2.75) -- ( 3.4,-0.25);
    \node at ( 4, 1.25) {$\ket{\psi}$};
    \foreach \i in {-3,...,3}
    {
        \draw[thick] (\i,-1) -- (\i,-6.5);
    }
    \draw[thick, fill=\gatecolor] (-3.25,-1.5) rectangle (-1.75,-2);
    \draw[thick, fill=\gatecolor] (-0.25,-1.5) rectangle ( 1.25,-2);
    \draw[thick, fill=\gatecolor] ( 1.75,-1.5) rectangle ( 2.25,-2);
    \draw[thick, fill=\gatecolor] (-2.25,-2.5) rectangle (-0.75,-3);
    \fill[\gatecolor]             ( 0.75,-2.5) rectangle ( 1.25,-3);
    \fill[\gatecolor]             ( 2.75,-2.5) rectangle ( 3.25,-3);
    \draw[thick] ( 1.25,-2.5) -- ( 0.75,-2.5) -- ( 0.75,-3) -- ( 1.25,-3);
    \draw[thick] ( 2.75,-2.5) -- ( 3.25,-2.5) -- ( 3.25,-3) -- ( 2.75,-3);
    \draw[thick, dashed] ( 1.25,-2.5) -- ( 2.75,-2.5);
    \draw[thick, dashed] ( 1.25,-3)   -- ( 2.75,-3);
    \draw[thick, fill=\gatecolor] (-1.25,-3.5) rectangle ( 0.25,-4);
    \draw[thick, fill=\gatecolor] ( 1.75,-3.5) rectangle ( 3.25,-4);
    \draw[thick, fill=\gatecolor] (-3.25,-3.5) rectangle (-2.75,-4);
    \fill[\gatecolor]             (-2.25,-4.5) rectangle (-1.75,-5);
    \fill[\gatecolor]             ( 0.75,-4.5) rectangle ( 1.25,-5);
    \draw[thick] (-1.75,-4.5) -- (-2.25,-4.5) -- (-2.25,-5) -- (-1.75,-5);
    \draw[thick] ( 0.75,-4.5) -- ( 1.25,-4.5) -- ( 1.25,-5) -- ( 0.75,-5);
    \draw[thick, dashed] (-1.75,-4.5) -- ( 0.75,-4.5);
    \draw[thick, dashed] (-1.75,-5)   -- ( 0.75,-5);
    \draw[thick, fill=\gatecolor] ( 2.75,-4.5) rectangle ( 3.25,-5);
    \draw[thick, fill=\gatecolor] (-3.25,-5.5) rectangle (-1.75,-6);
    \draw[thick, fill=\gatecolor] (-1.25,-5.5) rectangle ( 0.25,-6);
    \draw[thick, fill=\gatecolor] ( 1.75,-5.5) rectangle ( 2.25,-6);
    \draw[->] (-3.5,-3) -- (-3.5,-4.5);
    \node[rotate=-90] at (-3.65,-3.75) {time};
\end{tikzpicture}
\caption{Principal algorithmic paradigm in this work: an initial $N$-qubit quantum state $\ket{\psi}$ represented as tree tensor network, to which the gates of a quantum circuit are applied while preserving the tree structure.}
\label{fig:framework}
\end{figure}
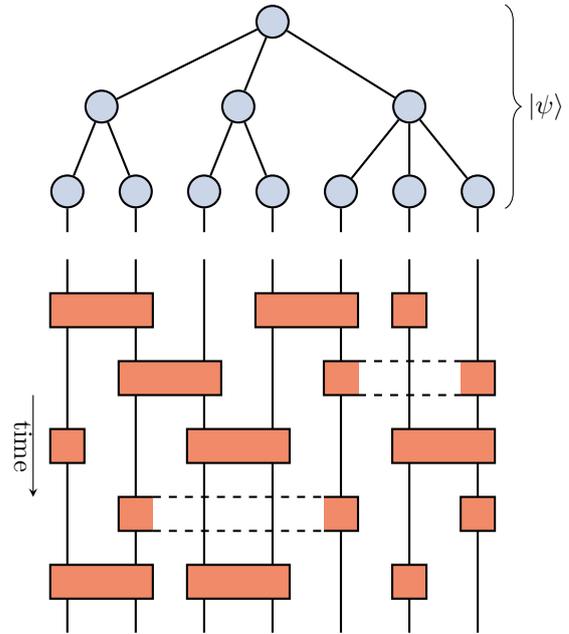

In general, TTNs share important advantages with MPSs~\cite{Ran2020}:
TTNs are efficiently contractable, incurring only polynomial cost on a classical computer for tensor nodes with bounded dimensions.
As a consequence, one can extract many quantities of physical interest from a quantum state in a TTN representation, for example, two-point (static) correlation functions, and one can draw unbiased (Born) probability samples from them~\cite{Cheng2019}.
We note, however, that the \textit{distance} between any pair of leaf nodes in a TTN, defined as the number of nodes traversed along the shortest path between the two chosen nodes, scales as $\mathcal{O}(\log(N))$, where $N$ denotes the overall number of lattice sites or qubits in the circuit. 
This compares favorably to the MPS representation, where the distance between leaf nodes scales as $\mathcal{O}(N)$.
Since connected correlation functions (i.e., covariances) typically decay exponentially with path length within a tensor network, TTNs can capture longer-range correlations as compared to MPS~\cite{Murg2010}. This property becomes relevant for quantum circuits with multi-qubit gates acting on ``distant'' qubits, or, in other words, in cases where one cannot devise a canonical linear ordering of qubits in which gates act solely in local neighborhoods. More concretely, we will identify scenarios where a tree representation provides a genuine advantage over an MPS in terms of the scaling of required internal bond dimensions, see Sect.~\ref{sec:comparison_mps} below.
In terms of tensor networks, reduced bond dimensions directly translate to improved contraction and simulation efficiency.

The method studied in this work can be classified as a statevector-based quantum circuit simulator and is summarized in Fig.~\ref{fig:framework}: a quantum state $\ket{\psi}$ (typically starting as a product state, e.g., $\ket{0}^{\otimes n}$) is represented as a TTN with an advantageous tree structure. The gates of the circuit are applied to $\ket{\psi}$ sequentially by absorbing them into the tree, as described in Sect.~\ref{sec:gate_application_procedure}.

\medskip

\paragraph*{Definitions and conventions.}

We consider qubits throughout to simplify the exposition, but the methods presented here are generalizable to ``qudits''.
We also assume that the circuit contains solely single- and two-qubit gates, with the remark that our algorithm can be extended to handle higher-order gates as well.

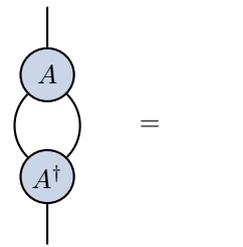
\begin{figure}
\centering
\begin{tikzpicture}[scale=0.9]
    \node[circle, draw, thick, minimum size=20, inner sep=0, fill=\nodecolor] (A)  at ( 0, 0.75) {$A$};
    \node[circle, draw, thick, minimum size=20, inner sep=0, fill=\nodecolor] (Ad) at ( 0,-0.75) {$A^{\dagger}$};
    \draw[thick] (A) to [out=-45, in=45]  (Ad);
    \draw[thick] (A) to [out=225, in=135] (Ad);
    \draw[thick] (A)  -- ( 0, 1.75);
    \draw[thick] (Ad) -- ( 0,-1.75);
    \node at ( 1.5, 0) {$=$};
    \draw[thick] ( 3,-1.75) -- ( 3, 1.75);
\end{tikzpicture}
\caption{Orthonormalization property of a tree node; the tree tensor network is of the canonical form if all nodes in the tree have this property.}
\label{fig:node_canonical}
\end{figure}

We say that a tree tensor network is of \emph{canonical form} (i.e., orthonormalized) if each node in the tree obeys the condition in Fig.~\ref{fig:node_canonical}, i.e., contracting a node tensor with its complex conjugate along its downstream legs gives the identity map.
Regarding the root node of the tree, one can formally attach a dummy upward leg with dimension $1$ to enforce this requirement.
With this property, a quantum state represented as a TTN is then normalized.

\section{Algorithm}
\label{sec:algorithm}

Our principal algorithm takes a quantum circuit description and a specification of the initial quantum state as inputs. The algorithm consists of two phases, as illustrated in Fig.~\ref{fig:framework}:
\begin{enumerate}
\item Expressing the initial quantum state as a TTN by mapping logical qubits to leaf nodes and identifying an advantageous rooted tree graph (i.e., connectivity of nodes) based on the given circuit.
\item Sequentially applying quantum circuit gates to the tree while preserving its graph structure, interleaved with re-orthonormalization of the tree tensor network.
\end{enumerate}

To ensure that the tree tensor network can be simulated on a classical computer, we require that each edge dimension is bounded by a chosen constant $D_{\max}$.
Naturally, this restricts the set of quantum circuits that can be simulated by our approach.
We will study how the $D_{\max}$ restriction is reflected in the feasible pattern of two-qubit gates in more detail in Sect.~\ref{sec:complexity_applicability}, and define the algorithm to determine a tree structure in the upcoming section.

\subsection{Tree structure search}
\label{sec:tree_structure}

In the first phase, the initial quantum state is converted to a TTN. Each logical qubit is mapped to a leaf node, which we group into subtrees, generating the tree-like structure.

Finding a truly optimal tree structure for a given problem depends on the problem statement \cite{Tagliacozzo2009}.
A similar technique in \cite{Szalay2015}  uses TTNs in a variational way to find ground states of given Hamiltonians.
They consider each node as a physical site, whereas in our case the physical sites live on the leaf nodes.
Instead of direct optimization, we employ a heuristic to find a tree topology that performs well in a variety of cases.
That being said, the topology optimizations discussed in \cite{Szalay2015, Murg2015} and our approach are similar in philosophy. 
In the case of \cite{Szalay2015}, they compute the mutual information and entanglement within different orbitals to arrive at a good tree architecture. 
In our case, the heuristic used to create the tree structure also looks at how the entanglement is distributed within the different physical parties, according to the gates to be applied. 
Thus our approach is an alternative to the mentioned ones for the case of circuit simulation. 

The heuristic is based on the insight that the maximal required dimension of internal edges can be upper-bounded a priori based on the to-be-applied circuit gates.
Single-qubit gates can be applied without increasing edge dimensions, hence, we focus solely on the effect of two-qubit gates.

\begin{figure}[!ht]
\centering
\begin{tikzpicture}[>=stealth, scale=0.9]
\draw[thick] (-1,-0.75) -- (-1, 0.75);
\draw[thick] ( 1,-0.75) -- ( 1, 0.75);
\draw[thick, fill=\gatecolor] (-1.25,-0.25) rectangle ( 1.25, 0.25);
\node at ( 0, 0) {\small $G$};
\draw[->] (0,-1.25) -- node[right]{\small SVD} (0,-1.75);
\begin{scope}[shift={(0,-3)}]
    \node[rectangle, thick, draw, fill=\gatecolor, minimum width=5mm, minimum height=5mm] (U) at (-1, 0) {\small $U$};
    \node[rectangle, thick, draw, fill=\gatecolor, minimum width=5mm, minimum height=5mm] (V) at ( 1, 0) {\small $V$};
    \node[circle, thick, draw, fill=\gatecolor, minimum size=5mm] (S) at ( 0, 0) {\small $S$};
    \draw[thick] (-1,-0.75) -- (U) -- (-1, 0.75);
    \draw[thick] ( 1,-0.75) -- (V) -- ( 1, 0.75);
    \draw[thick] (U) -- (S) -- (V);
    \draw[->] (0,-1.25) -- node[right]{\small absorb} (0,-1.75);
\end{scope}
\begin{scope}[shift={(0,-6)}]
    \node[rectangle, thick, draw, fill=\gatecolor, minimum width=5mm, minimum height=5mm] (E) at (-1, 0) {\small $G_l$};
    \node[rectangle, thick, draw, fill=\gatecolor, minimum width=5mm, minimum height=5mm] (F) at ( 1, 0) {\small $G_r$};
    \draw[thick] (-1,-0.75) -- (E) -- (-1, 0.75);
    \draw[thick] ( 1,-0.75) -- (F) -- ( 1, 0.75);
    \draw[thick] (E) -- node[below]{\small $\frac{1}{\sqrt{k}}$} (F);
\end{scope}
\end{tikzpicture}
\caption{Splitting of a two-qubit gate via SVD. The diagonal matrix $S$ contains the singular values. For the last step, one may absorb $\sqrt{S}$ both into $U$ and $V$. $k$ denotes the number of non-zero singular values, and the factor $1/\sqrt{k}$ serves as normalization when incorporating the connecting bond into the tree tensor network (see Fig.~\ref{fig:apply_gate} below).}
\label{fig:split_gate}
\end{figure}
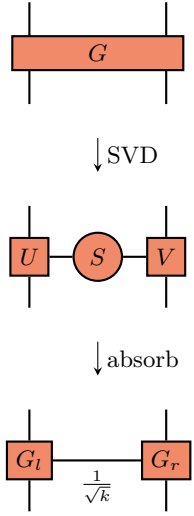

Similar to the case in an MPS, the application of a two-qubit gate increases the edge dimension between the corresponding leaf nodes and each intermediate node along the path that connects them, which we refer to as \emph{threading} the bond wire through the TTN.
As visualized in Fig.~\ref{fig:split_gate}, edge dimensions are increased by a factor of $k$, which can be determined by a singular value decomposition (SVD).
For the CNOT gate, $k = 2$, while in general $k = 4$, as in the case of the fSIM gate \cite{Arute2019}.

Thus, selecting subtrees of qubits that are highly entangled by the simulated circuit reduces the number of two-qubit gates spanning different subtrees, effectively increasing admissible circuit depth under the $D_{\max}$ restriction \cite{Jermyn2019}.
Additionally, threading connections through the tree incurs the overhead of renormalization of all nodes on the connecting path.

We identify such subtrees of qubits via the custom similarity function
\begin{equation}
\label{eq:distance_metric}
s_{\text{qc}}(q_i,q_j) = \abs{G(q_i) \cap G(q_j)} + \frac{1}{\abs{G(q_i)} + \abs{G(q_j)}}
\end{equation}
for $i \neq j$ and $\abs{G(q_i)} = \sharp{(\text{2-qubit gates})}\text{ on qubit } q_i$, where the first summand ranks qubit pairs with their entanglement and the second summand acts as a tie-breaker as it is $< 1$.
A generic clustering algorithm uses $s_{\text{qc}}$ to find similar-sized clusters over all qubits.
For simplicity, subtrees are constructed bottom-up for each identified cluster, putting all qubit pairs with the same similarity value under a common node.
Instead, one could recursively cluster all subtrees to a certain level and fall back to the bottom-up construction as a basis case.
Similar ideas have been explored in \cite{Ferrari2022, Szalay2015}.
\cite{Ferrari2022} construct adaptive-weighted binary TTNs bottom-up based on entanglement, which can be interpreted as a similarity function. 
One core difference is that our construction is not restricted to binary trees.
For quantum chemistry, the \emph{coordination number} of molecules can be used similarity function \cite{Szalay2015}.

One insight from early prototypes is the advantage of similar-sized subtrees under the root node, compared to skewed structures.
Favoring this pattern leads to similar subtree heights, and, consequently, ensures that the final tree structure is (almost) balanced. 
Thus, this is enforced in Algorithm~\ref{alg:tree_structure}, which shows the pseudocode for the first phase of our algorithm.

\begin{algorithm}
    \DontPrintSemicolon
    \SetKwProg{Fn}{Function}{:}{}
    \SetKw{KwTo}{in}
    \SetKwFunction{c}{cluster}\SetKwFunction{func}{find\_tree\_structure}\SetKwFunction{cs}{create\_subtree}\SetKwFunction{Node}{Node}\SetKwFunction{csm}{similarity\_matrix}\SetKwFunction{hca}{have\_common\_ancestor}\SetKwFunction{ac}{add\_connection}\SetKwFunction{atp}{add\_to\_parent}
    \SetKwData{sim}{similarity}\SetKwData{nc}{c}\SetKwData{cl}{cluster\_labels}\SetKwData{roots}{cluster\_roots}\SetKwData{lb}{label}\SetKwData{qc}{qc}
    \Fn{\func{\qc, \nc}}{
        \KwData{\qc $= Quantum\ Circuit,$ \nc$=Number\ of\ clusters$}
        \KwResult{Tree structure}
        \sim $\leftarrow$ \csm{\qc.gates}\tcp*{$s_{\text{qc}}$}
        \cl $\leftarrow$ \c{\sim, \nc}\; 
        \tcp*{any clustering, e.g., SpectralClustering}
        \roots $\leftarrow \emptyset$\;
        \ForEach{\lb \KwTo \cl }{
            \roots $\leftarrow$ \roots $\cup$ \cs{\qc{\lb}}\;
        }
        \tcc{$Node(children)$ creates a new subtree and builds the tree bottom up.} 
        \Return{\Node{\roots}}
    }
    \SetKwFunction{sort}{sort}\SetKwFunction{pair}{Pair}\SetKwFunction{sf}{similarity}
    \SetKwData{q}{qubits}\SetKwData{pairs}{pairs}\SetKwData{seen}{seen}\SetKwData{children}{children}\SetKwData{sim}{sim}
    \Fn{\cs{\q}}{
        \KwData{\q $=Qubits$}
        \KwResult{Cluster}
        \seen $\leftarrow \emptyset$\;
        \children $\leftarrow \emptyset$\;
        \pairs $\leftarrow$ \sort{\pair{\q}, $most\ similar$}\; 
        \sim $\leftarrow$ \sf{\pairs\hspace{-1mm}\rm{\texttt{[0][0]}}, \pairs\hspace{-1mm}\rm{\texttt{[0][1]}}}\;
        \tcp{\sf is based on Eq.~\eqref{eq:distance_metric}}
        \ForEach{$q_0, q_1$ \KwTo \pairs }{
            \uIf{$q_0\notin$ \seen}{
                \seen $\leftarrow$ \seen $\cup \ \{q_0\}$\;
                \children $\leftarrow$ \children $\cup \ \{q_0\}$\;
            }
            \uIf{$q_1\notin$ \seen}{
                \seen $\leftarrow$ \seen $\cup \ \{q_1\}$\;
                \children $\leftarrow$ \children $\cup \ \{q_1\}$\;
            }
            \uIf{\sim $>$ \sf{$q_0$, $q_1$}}{
                \children $\leftarrow \ \{$\Node{\children}$\}$\;
                \sim $\leftarrow$ \sf{$q_0$, $q_1$}\;
            }
        }
        \Return{\Node{\children}}
    }
    \caption{Tree structure search}
    \label{alg:tree_structure}
\end{algorithm}

\subsection{Gate application procedure}
\label{sec:gate_application_procedure}

Once the structure has been fixed, the tree tensor network has to represent the initial quantum state.
Typically this is a computational basis state, e.g., all qubits starting from $\ket{0}$.
For a computational basis state, one can set the dimensions of all internal edges to $1$ and initialize the internal tensors with the single entry $1$, while the two entries of a leaf tensor (of dimension $2 \times 1$) are the basis state entries of the corresponding qubit.
In other words, the initial tree can be regarded as a scaffolding to which gates are applied.

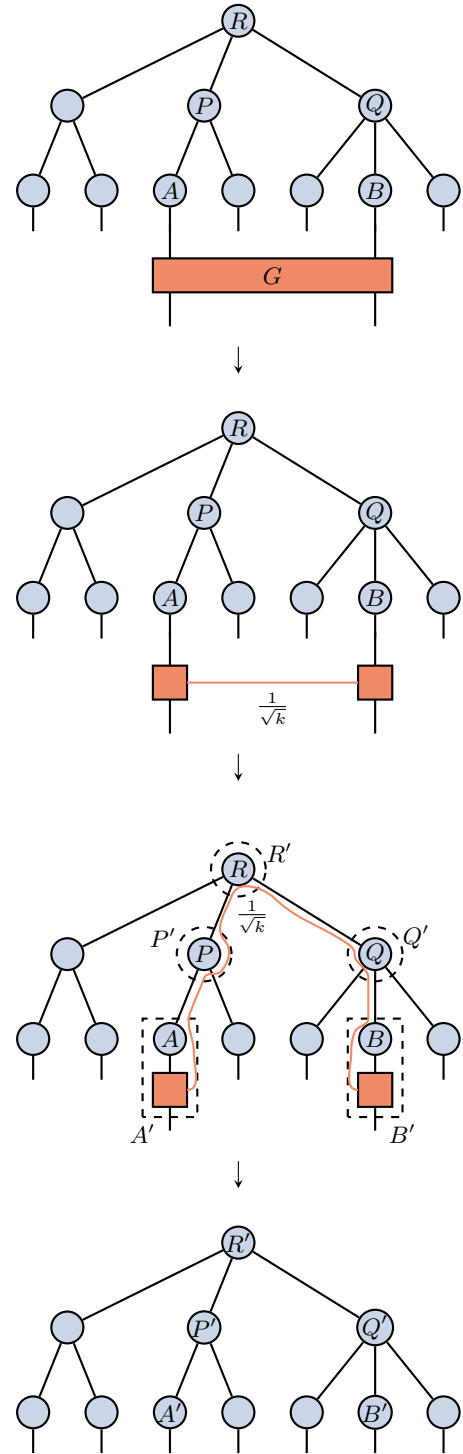
\begin{figure}[!ht]
\centering
\begin{tikzpicture}[>=stealth, scale=0.9]
    \node[circle, draw, thick, minimum size=12, inner sep=0, fill=\nodecolor] (C) at (-3,   0) {};
    \node[circle, draw, thick, minimum size=12, inner sep=0, fill=\nodecolor] (D) at (-2,   0) {};
    \node[circle, draw, thick, minimum size=12, inner sep=0, fill=\nodecolor] (A) at (-1,   0) {\small $A$};
    \node[circle, draw, thick, minimum size=12, inner sep=0, fill=\nodecolor] (E) at ( 0,   0) {};
    \node[circle, draw, thick, minimum size=12, inner sep=0, fill=\nodecolor] (F) at ( 1,   0) {};
    \node[circle, draw, thick, minimum size=12, inner sep=0, fill=\nodecolor] (B) at ( 2,   0) {\small $B$};
    \node[circle, draw, thick, minimum size=12, inner sep=0, fill=\nodecolor] (G) at ( 3,   0) {};
    \node[circle, draw, thick, minimum size=12, inner sep=0, fill=\nodecolor] (O) at (-2.5, 1.25) {};
    \node[circle, draw, thick, minimum size=12, inner sep=0, fill=\nodecolor] (P) at (-0.5, 1.25) {\small $P$};
    \node[circle, draw, thick, minimum size=12, inner sep=0, fill=\nodecolor] (Q) at ( 2,   1.25) {\small $Q$};
    \node[circle, draw, thick, minimum size=12, inner sep=0, fill=\nodecolor] (R) at ( 0,   2.5)  {\small $R$};
    \draw[thick] (-3,-0.6) -- (C);
    \draw[thick] (-2,-0.6) -- (D);
    \draw[thick] (-1,-0.6) -- (A);
    \draw[thick] ( 0,-0.6) -- (E);
    \draw[thick] ( 1,-0.6) -- (F);
    \draw[thick] ( 2,-0.6) -- (B);
    \draw[thick] ( 3,-0.6) -- (G);
    \draw[thick] (C) -- (O);
    \draw[thick] (D) -- (O);
    \draw[thick] (A) -- (P);
    \draw[thick] (E) -- (P);
    \draw[thick] (F) -- (Q);
    \draw[thick] (B) -- (Q);
    \draw[thick] (G) -- (Q);
    \draw[thick] (O) -- (R);
    \draw[thick] (P) -- (R);
    \draw[thick] (Q) -- (R);
    \draw[thick] (-1,-0.5) -- (-1,-2);
    \draw[thick] ( 2,-0.5) -- ( 2,-2);
    \draw[thick, fill=\gatecolor] (-1.25,-1) rectangle ( 2.25,-1.5);
    \node at ( 0.5,-1.25) {\small $G$};
    \draw[->] ( 0,-2.3) -- ( 0,-2.7);
    \begin{scope}[shift={( 0,-6)}]
    \node[circle, draw, thick, minimum size=12, inner sep=0, fill=\nodecolor] (C) at (-3,   0) {};
    \node[circle, draw, thick, minimum size=12, inner sep=0, fill=\nodecolor] (D) at (-2,   0) {};
    \node[circle, draw, thick, minimum size=12, inner sep=0, fill=\nodecolor] (A) at (-1,   0) {\small $A$};
    \node[circle, draw, thick, minimum size=12, inner sep=0, fill=\nodecolor] (E) at ( 0,   0) {};
    \node[circle, draw, thick, minimum size=12, inner sep=0, fill=\nodecolor] (F) at ( 1,   0) {};
    \node[circle, draw, thick, minimum size=12, inner sep=0, fill=\nodecolor] (B) at ( 2,   0) {\small $B$};
    \node[circle, draw, thick, minimum size=12, inner sep=0, fill=\nodecolor] (G) at ( 3,   0) {};
    \node[circle, draw, thick, minimum size=12, inner sep=0, fill=\nodecolor] (O) at (-2.5, 1.25) {};
    \node[circle, draw, thick, minimum size=12, inner sep=0, fill=\nodecolor] (P) at (-0.5, 1.25) {\small $P$};
    \node[circle, draw, thick, minimum size=12, inner sep=0, fill=\nodecolor] (Q) at ( 2,   1.25) {\small $Q$};
    \node[circle, draw, thick, minimum size=12, inner sep=0, fill=\nodecolor] (R) at ( 0,   2.5)  {\small $R$};
    \draw[thick] (-3,-0.6) -- (C);
    \draw[thick] (-2,-0.6) -- (D);
    \draw[thick] (-1,-0.6) -- (A);
    \draw[thick] ( 0,-0.6) -- (E);
    \draw[thick] ( 1,-0.6) -- (F);
    \draw[thick] ( 2,-0.6) -- (B);
    \draw[thick] ( 3,-0.6) -- (G);
    \draw[thick] (C) -- (O);
    \draw[thick] (D) -- (O);
    \draw[thick] (A) -- (P);
    \draw[thick] (E) -- (P);
    \draw[thick] (F) -- (Q);
    \draw[thick] (B) -- (Q);
    \draw[thick] (G) -- (Q);
    \draw[thick] (O) -- (R);
    \draw[thick] (P) -- (R);
    \draw[thick] (Q) -- (R);
    \draw[thick] (-1,-0.5) -- (-1,-2);
    \draw[thick] ( 2,-0.5) -- ( 2,-2);
    \draw[thick, fill=\gatecolor] (-1.25,-1) rectangle (-0.75,-1.5);
    \draw[thick, fill=\gatecolor] ( 1.75,-1) rectangle ( 2.25,-1.5);
    \draw[thick, \gatecolor] (-0.75,-1.25) --node[black, below] {\small $\frac{1}{\sqrt{k}}$} ( 1.75,-1.25);
    \draw[->] ( 0,-2.3) -- ( 0,-2.7);
    \end{scope}
    \begin{scope}[shift={( 0,-12.5)}]
    \node[circle, draw, thick, minimum size=12, inner sep=0, fill=\nodecolor] (C) at (-3,   0) {};
    \node[circle, draw, thick, minimum size=12, inner sep=0, fill=\nodecolor] (D) at (-2,   0) {};
    \node[circle, draw, thick, minimum size=12, inner sep=0, fill=\nodecolor] (A) at (-1,   0) {\small $A$};
    \node at (-1.4,-1.4) {\small $A'$};
    \node[circle, draw, thick, minimum size=12, inner sep=0, fill=\nodecolor] (E) at ( 0,   0) {};
    \node[circle, draw, thick, minimum size=12, inner sep=0, fill=\nodecolor] (F) at ( 1,   0) {};
    \node[circle, draw, thick, minimum size=12, inner sep=0, fill=\nodecolor] (B) at ( 2,   0) {\small $B$};
    \node at ( 2.4,-1.4) {\small $B'$};
    \node[circle, draw, thick, minimum size=12, inner sep=0, fill=\nodecolor] (G) at ( 3,   0) {};
    \node[circle, draw, thick, minimum size=12, inner sep=0, fill=\nodecolor] (O) at (-2.5, 1.25) {};
    \node[circle, draw, thick, minimum size=12, inner sep=0, fill=\nodecolor] (P) at (-0.5, 1.25) {\small $P$};
    \draw[thick, dashed] (P) circle (0.4);
    \node at (-1.1, 1.5) {\small $P'$};
    \node[circle, draw, thick, minimum size=12, inner sep=0, fill=\nodecolor] (Q) at ( 2,   1.25) {\small $Q$};
    \draw[thick, dashed] (Q) circle (0.4);
    \node at ( 2.6, 1.5) {\small $Q'$};
    \node[circle, draw, thick, minimum size=12, inner sep=0, fill=\nodecolor] (R) at ( 0,   2.5)  {\small $R$};
    \draw[thick, dashed] (R) circle (0.4);
    \node at ( 0.6, 2.75) {\small $R'$};
    \draw[thick] (-3,-0.6) -- (C);
    \draw[thick] (-2,-0.6) -- (D);
    \draw[thick] (-1,-1.35)-- (A);
    \draw[thick] ( 0,-0.6) -- (E);
    \draw[thick] ( 1,-0.6) -- (F);
    \draw[thick] ( 2,-1.35)-- (B);
    \draw[thick] ( 3,-0.6) -- (G);
    \draw[thick] (C) -- (O);
    \draw[thick] (D) -- (O);
    \draw[thick] (A) -- (P);
    \draw[thick] (E) -- (P);
    \draw[thick] (F) -- (Q);
    \draw[thick] (B) -- (Q);
    \draw[thick] (G) -- (Q);
    \draw[thick] (O) -- (R);
    \draw[thick] (P) -- (R);
    \draw[thick] (Q) -- (R);
    \draw[thick, fill=\gatecolor] (-1.25,-0.5) rectangle (-0.75,-1);
    \draw[thick, fill=\gatecolor] ( 1.75,-0.5) rectangle ( 2.25,-1);
    \draw[thick, dashed] (-1.4,-1.15) rectangle (-0.6, 0.3);
    \draw[thick, dashed] ( 1.6,-1.15) rectangle ( 2.4, 0.3);
    \draw[thick, \gatecolor] (-0.75, -0.75) to[out=0,in=-90] (-0.65, -0.2) to[out=90, in=-110] (-0.7, 0.4) to[out=70, in=-110] (-0.5, 0.9) to[out=70, in=-130] (-0.25, 1.0) to[out=50, in=-50] (-0.2, 1.45) to[out=130, in=-110] (-0.1, 2.1)  to[out=70, in=180] (0.1, 2.25) to[out=0, in=145] (0.5, 2.1) to[out=-45, in=145] (1.6, 1.4) to[out=-35, in=90] (1.7, 1.25) to[out=-90, in=90] (1.9, 0.9) to[out=-90, in=90] (1.9, 0.3) to[out=-90, in=70] (1.7, 0.1) to[out=-110, in=90] (1.65, -0.2) to[out=-90, in=180] (1.75, -0.75);
    \node at (0.2, 1.8) {\small $\frac{1}{\sqrt{k}}$};
    \draw[->] ( 0,-1.8) -- ( 0,-2.2);
    \end{scope}
    \begin{scope}[shift={( 0,-18)}]
    \node[circle, draw, thick, minimum size=12, inner sep=0, fill=\nodecolor] (C) at (-3,   0) {};
    \node[circle, draw, thick, minimum size=12, inner sep=0, fill=\nodecolor] (D) at (-2,   0) {};
    \node[circle, draw, thick, minimum size=12, inner sep=0, fill=\nodecolor] (A) at (-1,   0) {\small $A'$};
    \node[circle, draw, thick, minimum size=12, inner sep=0, fill=\nodecolor] (E) at ( 0,   0) {};
    \node[circle, draw, thick, minimum size=12, inner sep=0, fill=\nodecolor] (F) at ( 1,   0) {};
    \node[circle, draw, thick, minimum size=12, inner sep=0, fill=\nodecolor] (B) at ( 2,   0) {\small $B'$};
    \node[circle, draw, thick, minimum size=12, inner sep=0, fill=\nodecolor] (G) at ( 3,   0) {};
    \node[circle, draw, thick, minimum size=12, inner sep=0, fill=\nodecolor] (O) at (-2.5, 1.25) {};
    \node[circle, draw, thick, minimum size=12, inner sep=0, fill=\nodecolor] (P) at (-0.5, 1.25) {\small $P'$};
    \node[circle, draw, thick, minimum size=12, inner sep=0, fill=\nodecolor] (Q) at ( 2,   1.25) {\small $Q'$};
    \node[circle, draw, thick, minimum size=12, inner sep=0, fill=\nodecolor] (R) at ( 0,   2.5)  {\small $R'$};
    \draw[thick] (-3,-0.6) -- (C);
    \draw[thick] (-2,-0.6) -- (D);
    \draw[thick] (-1,-0.6) -- (A);
    \draw[thick] ( 0,-0.6) -- (E);
    \draw[thick] ( 1,-0.6) -- (F);
    \draw[thick] ( 2,-0.6) -- (B);
    \draw[thick] ( 3,-0.6) -- (G);
    \draw[thick] (C) -- (O);
    \draw[thick] (D) -- (O);
    \draw[thick] (A) -- (P);
    \draw[thick] (E) -- (P);
    \draw[thick] (F) -- (Q);
    \draw[thick] (B) -- (Q);
    \draw[thick] (G) -- (Q);
    \draw[thick] (O) -- (R);
    \draw[thick] (P) -- (R);
    \draw[thick] (Q) -- (R);
    \end{scope}
\end{tikzpicture}
\caption{Applying a two-qubit gate to a quantum state represented as a tree tensor network, preserving the tree structure and orthonormalization of the non-leaf nodes. The dashed circles and rectangles define the updated tensors by the enclosed items. }
\label{fig:apply_gate}
\end{figure}

\begin{figure}[!ht]
\centering
\begin{tikzpicture}[>=stealth, scale=0.9]
\node[circle, draw, thick, minimum size=20, inner sep=0, fill=\nodecolor] (A1)  at ( 0, 1) {$A$};
\node[circle, draw, thick, minimum size=20, inner sep=0, fill=\nodecolor] (A1d) at ( 0,-1) {$A^{\dagger}$};
\node at ( 0, 0.2) {\tiny $\frac{1}{\sqrt{k}}$};
\node at ( 0,-0.2) {\tiny $\frac{1}{\sqrt{k}}$};
\draw[thick, dashed] (A1)  circle (0.55);
\draw[thick, dashed] (A1d) circle (0.55);
\draw[thick] (A1) to [out=-45, in=45]  (A1d);
\draw[thick] (A1) to [out=225, in=135] (A1d);
\node at (-0.9, 1) {$A'$};
\node at (-0.9,-1) {$A'^{\dagger}$};
\draw[thick, \gatecolor] ( 0.5, 0) to[out= 90, in=0] ( 0, 0.55) to[out=180, in= 90] (-0.5, 0);
\draw[thick, \gatecolor] ( 0.5, 0) to[out=-90, in=0] ( 0,-0.55) to[out=180, in=-90] (-0.5, 0);
\draw[thick] (A1)  -- ( 0, 1.75);
\draw[thick] (A1d) -- ( 0,-1.75);
\node at ( 1.5, 0) {$=$};
\begin{scope}[shift={( 3, 0)}]
    \node[circle, draw, thick, minimum size=20, inner sep=0, fill=\nodecolor] (A2)  at ( 0, 1) {$A$};
    \node[circle, draw, thick, minimum size=20, inner sep=0, fill=\nodecolor] (A2d) at ( 0,-1) {$A^{\dagger}$};
    \draw[thick] (A2) to [out=-45, in=45]  (A2d);
    \draw[thick] (A2) to [out=225, in=135] (A2d);
    \draw[thick] (A2)  -- ( 0, 1.75);
    \draw[thick] (A2d) -- ( 0,-1.75);
    \draw[thick, \gatecolor] ( 1.5, 0) ellipse (0.25cm and 0.7cm);
    \node at ( 1.5, 0) {$\frac{1}{k}$};
    \node at ( 2.75, 0) {$=$};
\end{scope}
\draw[thick] ( 7,-1.75) -- ( 7, 1.75);
\begin{scope}[shift={( 0,-4)}]
\node[circle, draw, thick, minimum size=20, inner sep=0, fill=\nodecolor] (A1)  at ( 0, 1) {$A$};
\node[circle, draw, thick, minimum size=20, inner sep=0, fill=\nodecolor] (A1d) at ( 0,-1) {$A^{\dagger}$};
\draw[thick, dashed] (A1)  circle (0.55);
\draw[thick, dashed] (A1d) circle (0.55);
\node at (-0.9, 1) {$A'$};
\node at (-0.9,-1) {$A'^{\dagger}$};
\draw[thick] (A1) to [out=-45, in= 45] (A1d);
\draw[thick] (A1) to [out=225, in=135] (A1d);
\draw[thick] (A1)  -- ( 0, 1.75);
\draw[thick] (A1d) -- ( 0,-1.75);
\draw[thick, \gatecolor]                 ( 0.0824829, 1.75) -- (0.0824829, 1.46778);
\draw[thick, \gatecolor, line cap=round] ( 0.475, 1) arc (0: 80:0.475);
\draw[thick, \gatecolor, line cap=round] ( 0.475, 1) arc (0:-35:0.475);
\draw[thick, \gatecolor, line cap=round] ( 0.389097, 0.727551) to[out=-45, in=45] ( 0.389097,-0.727551);
\draw[thick, \gatecolor, line cap=round] ( 0.475,-1) arc (0:-80:0.475);
\draw[thick, \gatecolor, line cap=round] ( 0.475,-1) arc (0: 35:0.475);
\draw[thick, \gatecolor]                 ( 0.0824829,-1.46778) -- ( 0.0824829,-1.75);
\node at ( 1.5, 0) {$=$};
\begin{scope}[shift={( 3, 0)}]
    \node[circle, draw, thick, minimum size=20, inner sep=0, fill=\nodecolor] (A2)  at ( 0, 1) {$A$};
    \node[circle, draw, thick, minimum size=20, inner sep=0, fill=\nodecolor] (A2d) at ( 0,-1) {$A^{\dagger}$};
    \draw[thick] (A2) to [out=-45, in= 45] (A2d);
    \draw[thick] (A2) to [out=225, in=135] (A2d);
    \draw[thick] (A2)  -- ( 0, 1.75);
    \draw[thick] (A2d) -- ( 0,-1.75);
    \draw[thick, \gatecolor] ( 1.5,-1.75) -- ( 1.5, 1.75);
    \node at ( 2.75, 0) {$=$};
\end{scope}
\draw[thick] ( 7,-1.75) -- ( 7, 1.75);
\draw[thick, \gatecolor] ( 7.25,-1.75) -- ( 7.25, 1.75);
\end{scope}
\end{tikzpicture}
\caption{The canonical orthonormalization of a node tensor (see Fig.~\ref{fig:node_canonical}) is preserved when updating internal nodes of the tree during gate application (Fig.~\ref{fig:apply_gate}).}
\label{fig:node_orthonormality}
\end{figure}
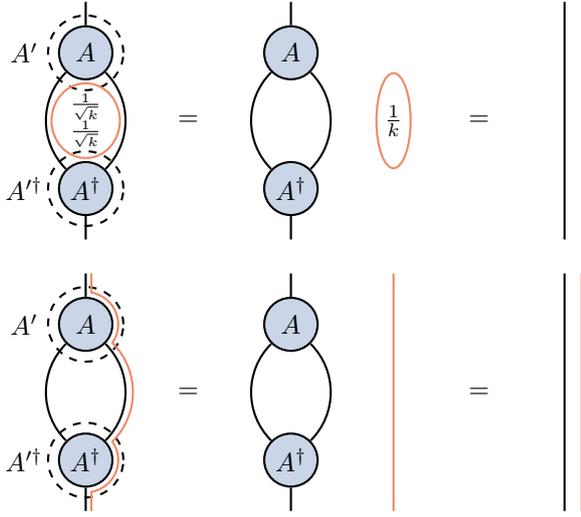

Fig.~\ref{fig:apply_gate} uses tensor diagram notation to illustrate incorporating (or ``absorbing'') a two-qubit gate into the tree, utilizing the decomposition step detailed in Fig.~\ref{fig:split_gate}.
After decomposing the gate, the connecting bond wire (red) is threaded through the tree structure.
The items enclosed in dotted lines define the updated tensors.
Note that the new non-leaf tensors ($P'$, $Q'$, $R'$ in the figure) essentially result from an outer product with the identity matrix, which can be recorded symbolically at first.
In other words, one may ``lazy-update" these tensors by delaying the operation until it becomes necessary.
The gate application procedure also preserves the orthonormality of non-leaf nodes -- this point is illustrated in Fig.~\ref{fig:node_orthonormality}.
Nevertheless, to return to canonical form, a re-orthonormalization sweep through the tree is necessary since the leaf nodes are no longer orthonormal.

\begin{figure}[!ht]
\centering
\begin{tikzpicture}[>=stealth, scale=0.9]
    \node[circle, draw, thick, minimum size=15, inner sep=0, fill=\nodecolor] (A) at ( 0,    0) {$A$};
    \node[circle, draw, thick, minimum size=15, inner sep=0, fill=\nodecolor] (P) at ( 2.25, 2.25) {$P$};
    \draw[thick, dash pattern=on 3pt off 3pt] (A) -- (0, -0.5);
    \draw[thick, dash pattern=on 3pt off 3pt, dash phase=3pt, \gatecolor] (A) -- (0, -0.5);
  
    \draw[thick, dash pattern=on 3pt off 3pt] (A) -- (P);
    \draw[thick, dash pattern=on 3pt off 3pt, dash phase=3pt, \gatecolor] (A) -- (P);
    \draw[thick, dash pattern=on 3pt off 3pt] (P) -- (2.25, 3);
    \draw[thick, dash pattern=on 3pt off 3pt, dash phase=3pt, \gatecolor] (P) -- (2.25, 3);    
    \draw[thick] (P) -- (2.75, 1.75);
    \draw[dashed] (-0.5, 0) -- (0.5, 0);
    \node at (1, 0) {\small SVD};
    \draw[->] ( 2.25 ,1) -- ( 3,1);
    \begin{scope}[shift={( 3,0)}]
        \node[circle, draw, thick, minimum size=15, inner sep=0, fill=\nodecolor] (A) at (0,   0) {$U$};
        \node[circle, draw, thick, minimum size=15, inner sep=0, fill=\nodecolor] (P) at (2.25,   2.25) {$P$};
        \node[circle, draw, thick, minimum size=15, inner sep=0, fill=\nodecolor] (B) at (0.75,   0.75) {$S$};
        \node[circle, draw, thick, minimum size=15, inner sep=0, fill=\nodecolor] (C) at (1.5,   1.5) {\small $V^{\dagger}$};
        \draw[thick, dash pattern=on 3pt off 3pt] (A) -- (0, -0.5);
        \draw[thick, dash pattern=on 3pt off 3pt, dash phase=3pt, \gatecolor] (A) -- (0, -0.5);
        \draw[thick, dash pattern=on 3pt off 3pt] (A) -- (B) -- (C) -- (P);
        \draw[thick, dash pattern=on 3pt off 3pt, dash phase=3pt, \gatecolor] (A) -- (B) -- (C) -- (P);
        \draw[thick, dash pattern=on 3pt off 3pt] (P) -- (2.25, 3);
        \draw[thick, dash pattern=on 3pt off 3pt, dash phase=3pt, \gatecolor] (P) -- (2.25, 3);    
        \draw[thick] (P) -- (2.75, 1.75);
        \draw[->] ( 2.25 ,1) -- ( 3,1);
    \end{scope}
    \begin{scope}[shift={( 6,0)}]
        \node[circle, draw, thick, minimum size=15, inner sep=0, fill=\nodecolor] (A) at (0,   0) {$A'$};
        \node[circle, draw, thick, minimum size=15, inner sep=0, fill=\nodecolor] (P) at (2.25,   2.25) {$P'$};
        \draw[thick, dash pattern=on 3pt off 3pt] (A) -- (0, -0.5);
        \draw[thick, dash pattern=on 3pt off 3pt, dash phase=3pt, \gatecolor] (A) -- (0, -0.5);
        \draw[thick, dash pattern=on 3pt off 3pt] (A) -- (P);
        \draw[thick, dash pattern=on 3pt off 3pt, dash phase=3pt, \gatecolor] (A) -- (P);
        \draw[thick, dash pattern=on 3pt off 3pt] (P) -- (2.25, 3);
        \draw[thick, dash pattern=on 3pt off 3pt, dash phase=3pt, \gatecolor] (P) -- (2.25, 3);
        \draw[thick] (P) -- (2.75, 1.75);
    \end{scope}
\end{tikzpicture}
\caption{Orthonormalization of a leaf $A$ in the tree via SVD: the $A$ tensor is replaced by the isometry $A'=U$ from the SVD, and the diagonal matrix of singular values, as well as $V^{\dagger}$, are absorbed in the parent tensor $P$.
The orthonormalization procedure is independent of gates as additional bonds can simply be bundled with the existing bonds indicated by the dashed line.
For intermediate nodes, the procedure stays the same, but additional wires need to be bundled as well.}
\label{fig:tree_orthonormalization_step}
\end{figure}
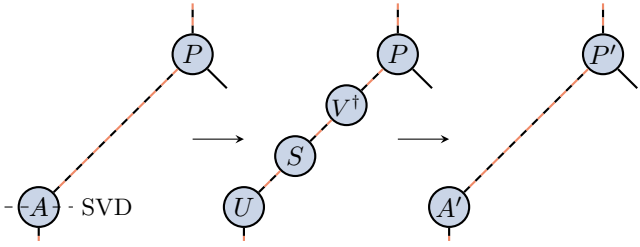

Re-orthonormalization proceeds from the leaf nodes upwards through the tree, using QR-decompositions or SVDs, as shown in Fig.~\ref{fig:tree_orthonormalization_step}. By using ``economical'' SVDs (explained later), the edge dimensions can only decrease or remain the same.

A variant of our algorithm is an approximation of the true output quantum state by retaining only the leading $D_{\max}$ singular values during the SVD-orthonormalization procedure.

\section{Tree tensor network capability and algorithmic analysis}
\label{sec:complexity_applicability}

In this section, we analyze the algorithmic complexity and identify the required properties of the quantum circuit such that it can still be simulated efficiently using our approach.
For specificity, our analysis refers to $m$-ary trees (see below), but note that the tree layout found by the tree structure search might result in more general architectures. These are mostly explored via numerical experiments in Sect.~\ref{sec:experiments}.

\subsection{Overall architecture and assumptions}

We impose the restriction that each edge dimension in the tree (i.e., dimension of any tensor leg) is bounded by a given number $D_{\max}$.
To simplify the analysis, let us assume a perfect $m$-ary tree, i.e., each non-leaf node has $m$ children, and that the tree is balanced.
For example, a binary tree corresponds to $m = 2$.
But note that the algorithm works for general rooted trees as well.
$D_{\max}$ should be chosen such that storing and working with such tensors is still possible given the available computational resources, memory, and hardware architecture.

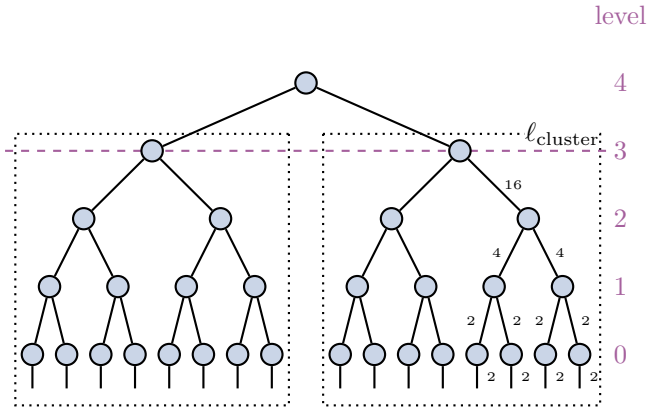
\begin{figure}[!ht]
\centering
\begin{tikzpicture}[>=stealth, scale=0.9]
    \draw[thick, dotted] ( 0.25,-0.75) rectangle ( 4.3, -4.75);
    \draw[thick, dotted] (-0.25,-0.75) rectangle (-4.25,-4.75);
    \node at (4.6,  0) {\color{mpurple} $4$};
    \node at (4.6, -1) {\color{mpurple} $3$};
    \node at (4.6, -2) {\color{mpurple} $2$};
    \node at (4.6, -3) {\color{mpurple} $1$};
    \node at (4.6, -4) {\color{mpurple} $0$};
    \node at (4.6,  1) {\color{mpurple} level};
    \fill[white] (3.2,-1) rectangle (4.25,-0.5);
    \node at (3.75,-0.75) {$\ell_{\text{cluster}}$};
    \draw[thick, dashed, color=mpurple] (-4.4,-1) -- (4.4,-1);

    \node[circle, draw, thick, minimum size=8, inner sep=0, fill=\nodecolor] (A)   at ( 0,    0) {};
    \node[circle, draw, thick, minimum size=8, inner sep=0, fill=\nodecolor] (B)   at (-2.25,-1) {};
    \node[circle, draw, thick, minimum size=8, inner sep=0, fill=\nodecolor] (C)   at ( 2.25,-1) {};
    \node[circle, draw, thick, minimum size=8, inner sep=0, fill=\nodecolor] (D)   at (-3.25,-2) {};
    \node[circle, draw, thick, minimum size=8, inner sep=0, fill=\nodecolor] (E)   at (-1.25,-2) {};
    \node[circle, draw, thick, minimum size=8, inner sep=0, fill=\nodecolor] (F)   at ( 1.25,-2) {};
    \node[circle, draw, thick, minimum size=8, inner sep=0, fill=\nodecolor] (G)   at ( 3.25,-2) {};
    \node[circle, draw, thick, minimum size=8, inner sep=0, fill=\nodecolor] (H)   at (-3.75,-3) {};
    \node[circle, draw, thick, minimum size=8, inner sep=0, fill=\nodecolor] (I)   at (-2.75,-3) {};
    \node[circle, draw, thick, minimum size=8, inner sep=0, fill=\nodecolor] (J)   at (-1.75,-3) {};
    \node[circle, draw, thick, minimum size=8, inner sep=0, fill=\nodecolor] (K)   at (-0.75,-3) {};
    \node[circle, draw, thick, minimum size=8, inner sep=0, fill=\nodecolor] (L)   at ( 0.75,-3) {};
    \node[circle, draw, thick, minimum size=8, inner sep=0, fill=\nodecolor] (M)   at ( 1.75,-3) {};
    \node[circle, draw, thick, minimum size=8, inner sep=0, fill=\nodecolor] (N)   at ( 2.75,-3) {};
    \node[circle, draw, thick, minimum size=8, inner sep=0, fill=\nodecolor] (O)   at ( 3.75,-3) {};
    \node[circle, draw, thick, minimum size=8, inner sep=0, fill=\nodecolor] (l0)  at (-4,   -4) {};
    \node[circle, draw, thick, minimum size=8, inner sep=0, fill=\nodecolor] (l1)  at (-3.5, -4) {};
    \node[circle, draw, thick, minimum size=8, inner sep=0, fill=\nodecolor] (l2)  at (-3,   -4) {};
    \node[circle, draw, thick, minimum size=8, inner sep=0, fill=\nodecolor] (l3)  at (-2.5, -4) {};
    \node[circle, draw, thick, minimum size=8, inner sep=0, fill=\nodecolor] (l4)  at (-2,   -4) {};
    \node[circle, draw, thick, minimum size=8, inner sep=0, fill=\nodecolor] (l5)  at (-1.5, -4) {};
    \node[circle, draw, thick, minimum size=8, inner sep=0, fill=\nodecolor] (l6)  at (-1,   -4) {};
    \node[circle, draw, thick, minimum size=8, inner sep=0, fill=\nodecolor] (l7)  at (-0.5, -4) {};
    \node[circle, draw, thick, minimum size=8, inner sep=0, fill=\nodecolor] (l8)  at ( 0.5, -4) {};
    \node[circle, draw, thick, minimum size=8, inner sep=0, fill=\nodecolor] (l9)  at ( 1,   -4) {};
    \node[circle, draw, thick, minimum size=8, inner sep=0, fill=\nodecolor] (l10) at ( 1.5, -4) {};
    \node[circle, draw, thick, minimum size=8, inner sep=0, fill=\nodecolor] (l11) at ( 2,   -4) {};
    \node[circle, draw, thick, minimum size=8, inner sep=0, fill=\nodecolor] (l12) at ( 2.5, -4) {};
    \node[circle, draw, thick, minimum size=8, inner sep=0, fill=\nodecolor] (l13) at ( 3,   -4) {};
    \node[circle, draw, thick, minimum size=8, inner sep=0, fill=\nodecolor] (l14) at ( 3.5, -4) {};
    \node[circle, draw, thick, minimum size=8, inner sep=0, fill=\nodecolor] (l15) at ( 4,   -4) {};

    \draw[thick] (A) -- (B);
    \draw[thick] (A) -- (C);
    \draw[thick] (B) -- (D);
    \draw[thick] (B) -- (E);
    \draw[thick] (C) -- (F);
    \draw[thick] (C) --node[right]  {\tiny 16} (G);
    \draw[thick] (D) -- (H);
    \draw[thick] (D) -- (I);
    \draw[thick] (E) -- (J);
    \draw[thick] (E) -- (K);
    \draw[thick] (F) -- (L);
    \draw[thick] (F) -- (M);
    \draw[thick] (G) --node[left]  {\tiny 4} (N);
    \draw[thick] (G) --node[right] {\tiny 4} (O);
    \draw[thick] (H) -- (l0);
    \draw[thick] (H) -- (l1);
    \draw[thick] (I) -- (l2);
    \draw[thick] (I) -- (l3);
    \draw[thick] (J) -- (l4);
    \draw[thick] (J) -- (l5);
    \draw[thick] (K) -- (l6);
    \draw[thick] (K) -- (l7);
    \draw[thick] (L) -- (l8);
    \draw[thick] (L) -- (l9);
    \draw[thick] (M) -- (l10);
    \draw[thick] (M) -- (l11);
    \draw[thick] (N) --node[left]  {\tiny 2} (l12);
    \draw[thick] (N) --node[right] {\tiny 2} (l13);
    \draw[thick] (O) --node[left]  {\tiny 2} (l14);
    \draw[thick] (O) --node[right] {\tiny 2} (l15);
    \draw[thick] (-4,  -4.5) -- (l0);
    \draw[thick] (-3.5,-4.5) -- (l1);
    \draw[thick] (-3,  -4.5) -- (l2);
    \draw[thick] (-2.5,-4.5) -- (l3);
    \draw[thick] (-2,  -4.5) -- (l4);
    \draw[thick] (-1.5,-4.5) -- (l5);
    \draw[thick] (-1,  -4.5) -- (l6);
    \draw[thick] (-0.5,-4.5) -- (l7);
    \draw[thick] ( 0.5,-4.5) -- (l8);
    \draw[thick] ( 1,  -4.5) -- (l9);
    \draw[thick] ( 1.5,-4.5) -- (l10);
    \draw[thick] ( 2,  -4.5) -- (l11);
    \draw[thick] ( 2.5,-4.5) --node[right] {\tiny 2} (l12);
    \draw[thick] ( 3,  -4.5) --node[right] {\tiny 2} (l13);
    \draw[thick] ( 3.5,-4.5) --node[right] {\tiny 2} (l14);
    \draw[thick] ( 4,  -4.5) --node[right] {\tiny 2} (l15);
\end{tikzpicture}
\caption{Maximally required edge dimensions up to level $3$ for a binary tree, see Eq.~\eqref{eq:child_dim_bound}. Assuming $D_{\max} = 16$, an arbitrary sequence of gates acting on the highlighted subtrees separately is feasible, while the number of two-qubit gates targeting one qubit from either subtree is restricted.}
\label{fig:tree_levels_dims}
\end{figure}

For the following, we first recall that an ``economical'' singular value decomposition of a matrix $A \in \C^{p \times q}$ is given by $A = U S V^{\dagger}$, with isometries $U \in \C^{p \times k}$ and $V \in \C^{q \times k}$ where $k = \min(p, q)$, and a diagonal matrix $S$ of singular values $\sigma_1 \ge \dots \ge \sigma_k \ge 0$.
In the context of Fig.~\ref{fig:tree_orthonormalization_step}, $p$ is the product of the dimensions of the downward-pointing child connections of $A$.
The term \emph{child connection} is used to describe an edge between nodes from level $\ell$ to $\ell-1$.
Assuming that these are all equal to some integer $D$ with $D \le D_{\max}$, $p = D^m$.
After the update, the connecting edge between $A$ and $P$ has dimension $k \le p = D^m$.
After an orthonormalization sweep through the tree starting from the leaf nodes, the dimensions of the child connections at level $\ell \ge 1$ are thus bounded by
\begin{equation}
\label{eq:child_dim_bound}
D_{\ell} \le 2^{(m^{\ell - 1})},
\end{equation}
as illustrated in Fig.~\ref{fig:tree_levels_dims}.
We count levels starting from $0$ at the leaf nodes.
The base $2$ in Eq.~\eqref{eq:child_dim_bound} stems from the dimension of a single qubit at the lowest level.

We define $\ell_{\text{cluster}}$ as the largest integer such that
\begin{equation}
2^{(m^{\ell_{\text{cluster}} - 1})} \le D_{\max}.
\end{equation}
According to Eq.~\eqref{eq:child_dim_bound}, the edge dimensions will never exceed $D_{\max}$ up to level $\ell_{\text{cluster}}$ in the tree.
This implies that any gate sequence \emph{within} a subtree of height $\ell_{\text{cluster}}$ is exact.
In the following, we denote these subtrees as \emph{clusters}.
A cluster thus contains $m^{\ell_{\text{cluster}}}$ qubits (assuming uniform height).
From another viewpoint, we can exactly represent any quantum state vector of the qubits in a cluster if it is unentangled with outside qubits.

Let us now consider the connecting edges above $\ell_{\text{cluster}}$ in the tree: we denote the bond dimension of a two-qubit gate $G$ by $k_G$, see Fig.~\ref{fig:split_gate}.
We say that a gate \emph{crosses} an edge $e$ if the gate bond threaded through the tree traverses $e$; for example, in Fig.~\ref{fig:apply_gate} the red curve crosses the edge between nodes $P$ and $R$.
The restriction imposed by $D_{\max}$ is thus certainly satisfied if
\begin{equation}
\label{eq:edge_gates_restriction}
\prod_{G \text{ crossing } e} k_G \le D_{\max}
\end{equation}
for any edge $e$ in the tree between nodes at or above level $\ell_{\text{cluster}}$.
In other words, the left term in Eq.~\eqref{eq:edge_gates_restriction} is an upper bound on the maximal entanglement acceptably generated by the gates crossing $e$.
We used Eq.~\eqref{eq:edge_gates_restriction} as motivation for the similarity function Eq.~\eqref{eq:distance_metric}. However, note that Eq.~\eqref{eq:edge_gates_restriction} does not consider possible simplifications in the contracted tensor structure, such as the (contrived) scenario when an arbitrary two-qubit gate $G$ is immediately followed by $G^{\dagger}$, which does not increase entanglement at all. Thus, Eq.~\eqref{eq:edge_gates_restriction} is a ``worst case'' bound on the applicable gates.

The level of the root node, $\ell_{\text{root}}$, denominates the overall height of the tree.
Thus, it can represent a quantum state with up to $N = m^{\ell_{\text{root}}}$ qubits.

\subsection{Computational cost and complexity analysis}

\subsubsection{Memory requirements}
\label{sec:memory_cost}

The number of nodes of the tree is less or equal to the node count of a perfect $m$-ary tree:
\begin{equation}
\sharp(\text{nodes}) \le \sum_{\ell=0}^{\ell_{\text{root}}} m^{\ell_{\text{root}} - \ell} = \frac{m^{\ell_{\text{root}} + 1} - 1}{m - 1}.
\end{equation}
As in Eq.~\eqref{eq:child_dim_bound}, we denote the largest appearing bond dimension of the child connections at level $\ell$ by $D_{\ell}$.
Thus a tensor at level $\ell$ has at most $D_{\ell}^m D_{\ell+1}$ entries: $m$ connections to its children and one connection to its parent.
Consequently, an upper bound on the overall to-be-stored (complex) numbers for the tree is
\begin{multline}
\label{eq:num_entries_bound_general}
\sharp(\text{entries})_{\text{tree}} \le m^{\ell_{\text{root}}} 2 D_1\\
+ \sum_{\ell=1}^{\ell_{\text{root}}-1} m^{\ell_{\text{root}}-\ell} D_{\ell}^m D_{\ell+1} + D_{\ell_{\text{root}}}^m.
\end{multline}
By construction $D_0 = 2$ (see Fig.~\ref{fig:tree_levels_dims}), for $1 \le \ell \le \ell_{\text{cluster}}$ the bound \eqref{eq:child_dim_bound} holds, and $D_{\ell} \le D_{\max}$ for $\ell > \ell_{\text{cluster}}$. Using Eq.~\eqref{eq:child_dim_bound} yields $D_{\ell}^m D_{\ell+1} \le 4^{(m^\ell)}$ for $1 \le \ell < \ell_{\text{cluster}}$. Inserting this bound into Eq.~\eqref{eq:num_entries_bound_general} leads to
\begin{multline}
\label{eq:num_entries_bound}
\sharp(\text{entries})_{\text{tree}} \le 4 \, m^{\ell_{\text{root}}} + \sum_{\ell=1}^{\ell_{\text{cluster}}-1} m^{\ell_{\text{root}}-\ell} \, 4^{(m^\ell)}\\
+ \sum_{\ell=\ell_{\text{cluster}}}^{\ell_{\text{root}}-1} m^{\ell_{\text{root}}-\ell} D_{\max}^{m+1} + D_{\max}^m.
\end{multline}
To arrive at a compact formula, we can use that $D_{\ell} \le D_{\max}$, and thus
\begin{equation}
\sharp(\text{entries})_{\text{tree}} \le \sharp(\text{nodes}) \, D_{\max}^{m+1} \le \frac{m N - 1}{m - 1} D_{\max}^{m+1}.
\end{equation}
Note that this bound is not reached because the root can only have up to $D_{\max}^m$ entries, and the leaf nodes have no children.
In summary, $\sharp(\text{entries})_{\text{tree}}$ grows linearly with qubit count $N$ (for fixed $m$ and $D_{\max}$).

\subsubsection{Gate application and orthonormalization}

The orthonormalization sweep through the tree is the most computationally expensive step in the gate application procedure. In the worst case, the bond wire threaded through the tree visits all levels up to the root node. Thus $2 \ell_{\text{root}}$ tensors along the path need to be re-orthonormalized. Recalling that the SVD decomposition of a $p \times q$ matrix costs $\mathcal{O}(\min(p q^2, p^2 q))$ floating-point operations, we arrive at the overall cost
\begin{equation}
\sharp(\text{FLOPS})_{\text{gate}} \le \mathcal{O}\!\left(\ell_{\text{root}} \cdot D_{\max}^{m+2}\right) = \mathcal{O}\!\left(\log_m(N) \cdot D_{\max}^{m + 2}\right)
\end{equation}
for an orthonormalization sweep after applying a gate.

\subsection{Admissible gate patterns}

Finally, we identify admissible gate patterns such that Eq.~\eqref{eq:edge_gates_restriction} certainly holds. To simplify the analysis, we do not take the option for truncation (omitting small singular values during orthonormalization) into account here, which could facilitate the application of additional gates while respecting the $D_{\max}$ restriction. Also, for simplicity we throughout set $k_G = 4$, the largest possible bond dimension of a two-qubit gate.

As described above, applying gates on qubits within the same cluster is unrestrained, so we only need to consider cases where the two qubits lie within different clusters.

\begin{figure}[!ht]
\centering
\def\cosfourty{0.766044}
\def\sinfourty{0.642788}
\subfloat[$m = 3$, $D_{\max} = 16$]{%
\begin{tikzpicture}[scale=0.9]
    \draw[thick, \gatecolor] ( 0.1, 2.5) -- ( 0.1, 0.5) to[out=-90, in=140] ( 0.5*\cosfourty + 0.1*\sinfourty, -0.5*\sinfourty + 0.1*\cosfourty) -- ( 2.5*\cosfourty + 0.1*\sinfourty, -2.5*\sinfourty + 0.1*\cosfourty);
    \draw[thick, \gatecolor] (-0.1, 2.5) -- (-0.1, 0.5) to[out=-90, in= 40] (-0.5*\cosfourty - 0.1*\sinfourty, -0.5*\sinfourty + 0.1*\cosfourty) -- (-2.5*\cosfourty - 0.1*\sinfourty, -2.5*\sinfourty + 0.1*\cosfourty);
    \draw[thick, \gatecolor] ( 0.1,-2.0) -- ( 0.1,-0.5) to[out= 90, in=140] ( 0.5*\cosfourty - 0.1*\sinfourty, -0.5*\sinfourty - 0.1*\cosfourty) -- ( 2.5*\cosfourty - 0.1*\sinfourty, -2.5*\sinfourty - 0.1*\cosfourty);
    \draw[thick, \gatecolor] (-0.1,-2.0) -- (-0.1,-0.5) to[out= 90, in= 40] (-0.5*\cosfourty + 0.1*\sinfourty, -0.5*\sinfourty - 0.1*\cosfourty) -- (-2.5*\cosfourty + 0.1*\sinfourty, -2.5*\sinfourty - 0.1*\cosfourty);
    \node[circle, draw, thick, minimum size=30, inner sep=0] at (0,0) {};
\end{tikzpicture}}%
\hspace{1cm}%
\subfloat[$m = 3$, $D_{\max} = 64$]{%
\begin{tikzpicture}[scale=0.9]
    \draw[thick, \gatecolor] ( 0,   2.5) -- ( 0,  -2.0);
    \draw[thick, \gatecolor] ( 0.1, 2.5) -- ( 0.1, 0.5) to[out=-90, in=140] ( 0.5*\cosfourty + 0.1*\sinfourty, -0.5*\sinfourty + 0.1*\cosfourty) -- ( 2.5*\cosfourty + 0.1*\sinfourty, -2.5*\sinfourty + 0.1*\cosfourty);
    \draw[thick, \gatecolor] (-0.1, 2.5) -- (-0.1, 0.5) to[out=-90, in= 40] (-0.5*\cosfourty - 0.1*\sinfourty, -0.5*\sinfourty + 0.1*\cosfourty) -- (-2.5*\cosfourty - 0.1*\sinfourty, -2.5*\sinfourty + 0.1*\cosfourty);
    \draw[thick, \gatecolor] ( 0.1,-2.0) -- ( 0.1,-0.5) to[out= 90, in=140] ( 0.5*\cosfourty - 0.1*\sinfourty, -0.5*\sinfourty - 0.1*\cosfourty) -- ( 2.5*\cosfourty - 0.1*\sinfourty, -2.5*\sinfourty - 0.1*\cosfourty);
    \draw[thick, \gatecolor] (-0.1,-2.0) -- (-0.1,-0.5) to[out= 90, in= 40] (-0.5*\cosfourty + 0.1*\sinfourty, -0.5*\sinfourty - 0.1*\cosfourty) -- (-2.5*\cosfourty + 0.1*\sinfourty, -2.5*\sinfourty - 0.1*\cosfourty);
    \draw[thick, \gatecolor] (-2.5*\cosfourty, -2.5*\sinfourty) -- (-0.5*\cosfourty, -0.5*\sinfourty) to[out=40, in=140] ( 0.5*\cosfourty, -0.5*\sinfourty) -- ( 2.5*\cosfourty, -2.5*\sinfourty);
    \node[circle, draw, thick, minimum size=30, inner sep=0] at (0,0) {};
\end{tikzpicture}}
\caption{Examples of admissible threading of bond wires through a node with $m = 3$ children, such that the dimension of each edge remains bounded by $D_{\max}$. The subtrees can be connected pairwise by a gate, as well as upstream with other (more distant) qubits.}
\label{fig:admissible_gate_pattern}
\end{figure}
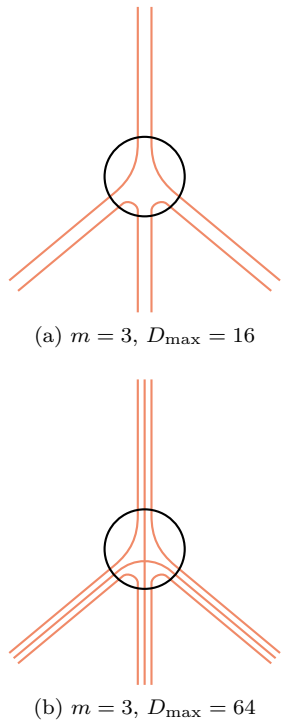

Solving Eq.~\eqref{eq:edge_gates_restriction} for the number of gates crossing an edge $e$ gives
\begin{equation}
\sharp(\text{gates crossing } e) \le \log_4(D_{\max}) = \frac{1}{2} \log_2(D_{\max}).
\end{equation}
If the bond wire of a gate traverses a node above level $\ell_{\text{cluster}}$, it multiplies the dimensions of two of its legs by the factor $4$ (without truncation), i.e., a factor of $16$ more entries is now required for this node. This gives an upper bound on the number of gates with bond wires crossing a given node $A$:
\begin{equation}
\sharp(\text{gates crossing } A) \le \frac{1}{4} \log_2(D_{\max}^{m+1}) = \frac{m+1}{4} \log_2(D_{\max}).
\end{equation}
Fig.~\ref{fig:admissible_gate_pattern} shows admissible scenarios for $m = 3$ and $D_{\max} = 16$ and $D_{\max} = 64$, respectively.

\begin{figure}
\def\tricoordA{{ 0.866025, -0.5}}
\def\tricoordB{{ 0,         1  }}
\def\tricoordC{{-0.866025, -0.5}}
\def\tricoords{{\tricoordA}, {\tricoordB}, {\tricoordC}}
\centering
\subfloat[$m = 3$, $D_{\max} = 16$]{\label{fig:example_connectivity_Dmax16} %
\begin{tikzpicture}[>=stealth, scale=0.9]
\draw[very thin, fill=gray!10, rounded corners=14] (5.1*\tricoordA[0], 5.1*\tricoordA[1]) -- (5.1*\tricoordB[0], 5.1*\tricoordB[1]) -- (5.1*\tricoordC[0], 5.1*\tricoordC[1]) -- cycle;
\foreach \c [count = \i] in \tricoords {
    \begin{scope}[shift={(2.5*\c[0], 2.5*\c[1])}]
    \draw[very thin, fill=gray!25, rounded corners=9] (2.25*\tricoordA[0], 2.25*\tricoordA[1]) -- (2.25*\tricoordB[0], 2.25*\tricoordB[1]) -- (2.25*\tricoordC[0], 2.25*\tricoordC[1]) -- cycle;
    \foreach \d [count = \j] in \tricoords {
        \begin{scope}[shift={(\d[0], \d[1])}]
        \draw[very thin, fill=gray!40, rounded corners=3] (0.9*\tricoordA[0], 0.9*\tricoordA[1]) -- (0.9*\tricoordB[0], 0.9*\tricoordB[1]) -- (0.9*\tricoordC[0], 0.9*\tricoordC[1]) -- cycle;
        \foreach \e [count = \k] in \tricoords {
            \begin{scope}[shift={(0.4*\e[0], 0.4*\e[1])}]
                \foreach \f [count = \l] in \tricoords {
                \begin{scope}[shift={(0.2*\f[0], 0.2*\f[1])}]
                \node[circle, draw, thick, minimum size=4, inner sep=0, fill=\nodecolor] (Q\i\j\k\l) at (0, 0) {};
                \end{scope}
            }
            \end{scope}
        }
        \draw[thick] (Q\i\j11) -- (Q\i\j12) -- (Q\i\j21) -- (Q\i\j22) -- (Q\i\j23) -- (Q\i\j32) -- (Q\i\j33) -- (Q\i\j31) -- (Q\i\j13) -- (Q\i\j11);
        \draw[thick] (Q\i\j12) -- (Q\i\j13);
        \draw[thick] (Q\i\j23) -- (Q\i\j21);
        \draw[thick] (Q\i\j31) -- (Q\i\j32);
        \draw[thick] (Q\i\j32) -- (Q\i\j11) -- (Q\i\j23);
        \draw[thick] (Q\i\j13) -- (Q\i\j22) -- (Q\i\j31);
        \draw[thick] (Q\i\j21) -- (Q\i\j33) -- (Q\i\j12);
        \end{scope}
    }
    \end{scope}
}
\draw[thick] (Q1311) -- (Q1133);
\draw[thick] (Q1122) -- (Q1213);

\draw[thick] (Q2322) -- (Q2133);
\draw[thick] (Q2231) -- (Q2312);

\draw[thick] (Q3122) -- (Q3211);
\draw[thick] (Q3231) -- (Q3321);

\draw[thick] (Q1222) -- (Q2113);
\draw[thick] (Q3112) -- (Q1332);

\draw[thick] (Q2222) -- (5.2*\tricoordB[0], 5.2*\tricoordB[1]);
\draw[thick] (Q3333) -- (5.2*\tricoordC[0], 5.2*\tricoordC[1]);
\draw[thick, dotted] (5.2*\tricoordB[0], 5.2*\tricoordB[1]) -- (5.5*\tricoordB[0], 5.5*\tricoordB[1]);
\draw[thick, dotted] (5.2*\tricoordC[0], 5.2*\tricoordC[1]) -- (5.5*\tricoordC[0], 5.5*\tricoordC[1]);
\end{tikzpicture}} \\
\subfloat[$m = 3$, $D_{\max} = 64$]{\label{fig:example_connectivity_Dmax64} %
\begin{tikzpicture}[>=stealth, scale=0.9]
\draw[very thin, fill=gray!10, rounded corners=14] (5.1*\tricoordA[0], 5.1*\tricoordA[1]) -- (5.1*\tricoordB[0], 5.1*\tricoordB[1]) -- (5.1*\tricoordC[0], 5.1*\tricoordC[1]) -- cycle;
\foreach \c [count = \i] in \tricoords {
    \begin{scope}[shift={(2.5*\c[0], 2.5*\c[1])}]
    \draw[very thin, fill=gray!25, rounded corners=9] (2.25*\tricoordA[0], 2.25*\tricoordA[1]) -- (2.25*\tricoordB[0], 2.25*\tricoordB[1]) -- (2.25*\tricoordC[0], 2.25*\tricoordC[1]) -- cycle;
    \foreach \d [count = \j] in \tricoords {
        \begin{scope}[shift={(\d[0], \d[1])}]
        \draw[very thin, fill=gray!40, rounded corners=3] (0.9*\tricoordA[0], 0.9*\tricoordA[1]) -- (0.9*\tricoordB[0], 0.9*\tricoordB[1]) -- (0.9*\tricoordC[0], 0.9*\tricoordC[1]) -- cycle;
        \foreach \e [count = \k] in \tricoords {
            \begin{scope}[shift={(0.4*\e[0], 0.4*\e[1])}]
                \foreach \f [count = \l] in \tricoords {
                \begin{scope}[shift={(0.2*\f[0], 0.2*\f[1])}]
                \node[circle, draw, thick, minimum size=4, inner sep=0, fill=\nodecolor] (Q\i\j\k\l) at (0, 0) {};
                \end{scope}
            }
            \end{scope}
        }
        \draw[thick] (Q\i\j11) -- (Q\i\j12) -- (Q\i\j21) -- (Q\i\j22) -- (Q\i\j23) -- (Q\i\j32) -- (Q\i\j33) -- (Q\i\j31) -- (Q\i\j13) -- (Q\i\j11);
        \draw[thick] (Q\i\j12) -- (Q\i\j13);
        \draw[thick] (Q\i\j23) -- (Q\i\j21);
        \draw[thick] (Q\i\j31) -- (Q\i\j32);
        \draw[thick] (Q\i\j32) -- (Q\i\j11) -- (Q\i\j23);
        \draw[thick] (Q\i\j13) -- (Q\i\j22) -- (Q\i\j31);
        \draw[thick] (Q\i\j21) -- (Q\i\j33) -- (Q\i\j12);
        \end{scope}
    }
    \end{scope}
}
\draw[thick] (Q1311) -- (Q1133);
\draw[thick] (Q1122) -- (Q1213);
\draw[thick] (Q1231) -- (Q1321);

\draw[thick] (Q2322) -- (Q2133);
\draw[thick] (Q2123) -- (Q2211);
\draw[thick] (Q2231) -- (Q2312);

\draw[thick] (Q3311) -- (Q3132);
\draw[thick] (Q3122) -- (Q3211);
\draw[thick] (Q3231) -- (Q3321);

\draw[thick] (Q1222) -- (Q2113);
\draw[thick] (Q3112) -- (Q1332);
\draw[thick] (Q2333) -- (Q3222);

\draw[thick] (Q1111) -- (5.2*\tricoordA[0], 5.2*\tricoordA[1]);
\draw[thick] (Q2222) -- (5.2*\tricoordB[0], 5.2*\tricoordB[1]);
\draw[thick] (Q3333) -- (5.2*\tricoordC[0], 5.2*\tricoordC[1]);
\draw[thick, dotted] (5.2*\tricoordA[0], 5.2*\tricoordA[1]) -- (5.5*\tricoordA[0], 5.5*\tricoordA[1]);
\draw[thick, dotted] (5.2*\tricoordB[0], 5.2*\tricoordB[1]) -- (5.5*\tricoordB[0], 5.5*\tricoordB[1]);
\draw[thick, dotted] (5.2*\tricoordC[0], 5.2*\tricoordC[1]) -- (5.5*\tricoordC[0], 5.5*\tricoordC[1]);
\end{tikzpicture}
}
\caption{%
Top-down view of an admissible quantum gate pattern such that a perfect $3$-ary tree representation of the output quantum state is feasible.
The circles denote physical qubits, corresponding to leaf nodes of the tree, and the thick lines two-qubit gates (ordering in time not relevant).
The nested triangles indicate a partitioning into subtrees, using nodes of the form shown in Fig.~\ref{fig:admissible_gate_pattern}.
An arbitrary number of quantum gates can act within the smallest triangles (clusters with $9$ qubits), corresponding to subtrees at level $\ell_{\text{cluster}} = 2$.}
\label{fig:example_connectivity}
\end{figure}
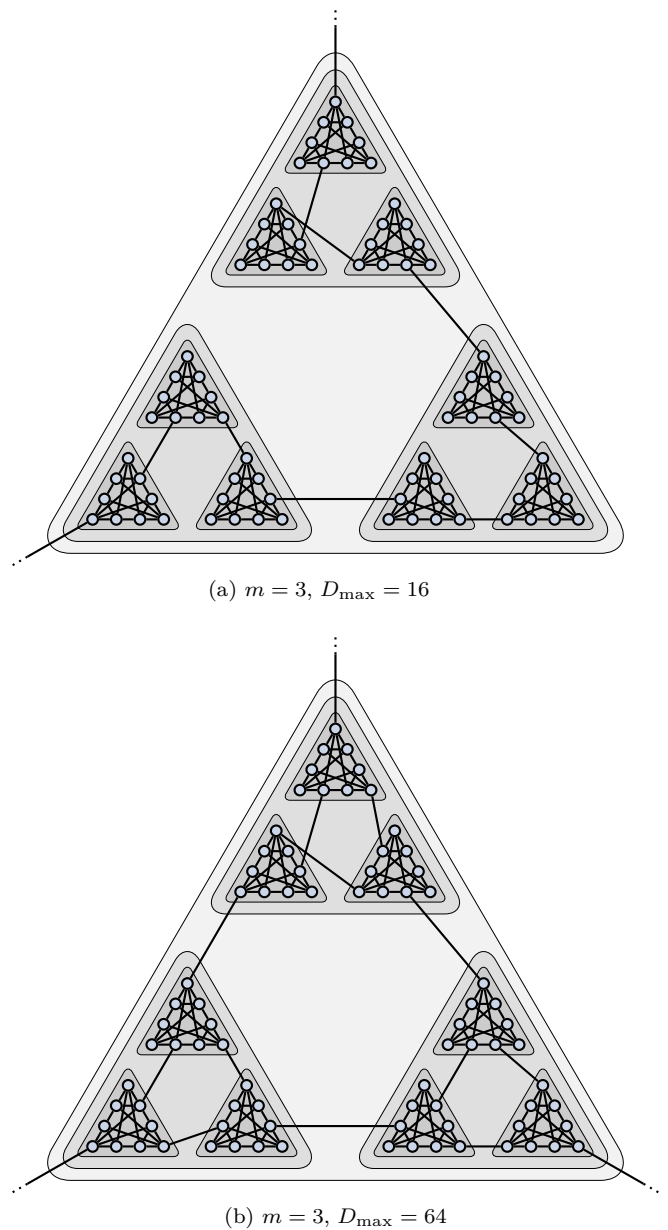

This pattern can be applied recursively to construct a tree.
Fig.~\ref{fig:example_connectivity} shows examples of the corresponding qubit connectivity and nested triangles defining the nodes on higher levels of the tree.
For $m = 3$ and $D_{\max} = 16, 64$, one obtains $\ell_{\text{cluster}} = 2$, such that each cluster contains nine qubits.
In the scenario of Fig.~\ref{fig:example_connectivity_Dmax16}, the inter-cluster bonds form a linear path visiting all clusters sequentially, whereas this is not the case in Fig.~\ref{fig:example_connectivity_Dmax64}.

\subsection{Comparison with matrix product states}
\label{sec:comparison_mps}

\begin{figure}
\def\tricoordA{{ 0.866025, -0.5}}
\def\tricoordB{{ 0,         1  }}
\def\tricoordC{{-0.866025, -0.5}}
\centering
\begin{tikzpicture}[>=stealth, scale=0.9]
\begin{scope}[shift={(-2.5,0)}]
    \node[circle, draw, thick, minimum size=12, inner sep=0] (A1) at (\tricoordA[0], \tricoordA[1]) {$1$};
    \node[circle, draw, thick, minimum size=12, inner sep=0] (A2) at (\tricoordB[0], \tricoordB[1]) {$2$};
    \node[circle, draw, thick, minimum size=12, inner sep=0] (A3) at (\tricoordC[0], \tricoordC[1]) {$3$};
    \draw[thick] (A1) -- (A2) -- (A3) -- (A1);
\end{scope}
\node at (0,0) {$\leadsto$};
\begin{scope}[shift={( 2.5,0)}]
    \draw[thick] (\tricoordA[0] + 0.1*\tricoordC[0], \tricoordA[1] + 0.1*\tricoordC[1]) -- (0.25*\tricoordA[0] + 0.75*\tricoordB[0] + 0.1*\tricoordC[0], 0.25*\tricoordA[1] + 0.75*\tricoordB[1] + 0.1*\tricoordC[1]) to[out=120, in=60] (0.25*\tricoordC[0] - 0.75*\tricoordB[0] + 0.1*\tricoordA[0], 0.25*\tricoordC[1] + 0.75*\tricoordB[1] + 0.1*\tricoordA[1]) -- (\tricoordC[0] + 0.1*\tricoordA[0], \tricoordC[1] + 0.1*\tricoordA[1]);
    \node[circle, draw, thick, fill=white, minimum size=12, inner sep=0] (B1) at (\tricoordA[0], \tricoordA[1]) {$1$};
    \node[circle, draw, thick, fill=white, minimum size=12, inner sep=0] (B2) at (\tricoordB[0], \tricoordB[1]) {$2$};
    \node[circle, draw, thick, fill=white, minimum size=12, inner sep=0] (B3) at (\tricoordC[0], \tricoordC[1]) {$3$};
    \draw[thick] (B1) -- (B2) -- (B3);
\end{scope}
\end{tikzpicture}
\caption{Pairwise connections between three qubits (or sites) require threading of one connection through an intermediate tensor in an MPS representation.}
\label{fig:mps_bond_routing}
\end{figure}
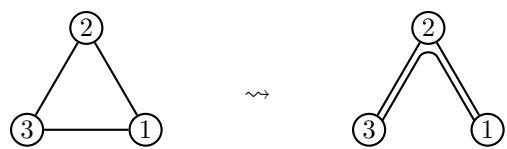

In this section, we argue that the tree representation can provide a genuine asymptotic advantage compared to an MPS:
for certain circuits, the required virtual bond dimension of an MPS will diverge as the number of qubits $N \to \infty$, while the dimensions of a TTN remain bounded. 
To explain the argument, consider the connectivity pattern in Fig.~\ref{fig:example_connectivity_Dmax64}.
First, note that an MPS imposes a linear ordering of the qubits and that a two-qubit gate acting on qubits $i$ and $k$ ($i < k$) increases the virtual bonds of all intermediate MPS tensors (sites $j$ with $i < j < k$).
Fig.~\ref{fig:mps_bond_routing} provides a simplified illustration of this point: the bond connecting $1$ with $3$ needs to be threaded through $2$ in an MPS representation.
For the connectivity pattern in Fig.~\ref{fig:example_connectivity_Dmax64}, and given an arbitrary ordering of the qubits, there exists a qubit playing the role of site $2$ at each level within every triangle.
Since this observation can be applied recursively to all levels, at least one of the virtual bond dimensions grows as $4^{\ell}$, with $\ell$ the number of levels.
In contrast, using the TTN approach, a uniformly bounded bond dimension suffices.

Applied to the simulation this has the following effect.
For the individual tree nodes, the cost of updating might be more expensive compared to the MPS.
But this difference is offset by the bound dimensionality.
The large cost of a full update is amortized by the savings of the common updates. 
More freely speaking, the necessity of a very expensive update is scarce in the TTN.
In an MPS, the still expensive updates are common enough, to produce a disadvantage.

\section{Numerical experiments}
\label{sec:experiments}

To complement the theoretical analysis in Sect.~\ref{sec:complexity_applicability}, we run and analyze numerical quantum circuit simulations based on the tree representation for the statevector, and compare them with a basic MPS-based simulator \footnote{Our implementation for the TTN and MPS simulators is available at \url{https://github.com/Gistbatch/tree-tensor-network-simulator}}.
For our study, one-qubit gates are not considered as they do not impact the dimensionality, and the bond dimension of each gate is assumed to be $k_G = 4$ unless stated otherwise.

All experiments are performed on a CPU (AMD Ryzen 7 3700) with 32 GB of RAM.
We explored both SVDs and QR decompositions for orthonormalization procedures, with comparable performance; the shown data uses SVDs throughout.
To test our code, we simulate a locally interacting circuit on an $n\times n$ lattice (as in Fig.~\ref{fig:sycamore}) with a depth of eight, based on Google Sycamore experiment \cite{Arute2019}.
This circuit will be referred to as a \emph{lattice} circuit. We also design a \emph{tree-like} circuit (as in Fig.~\ref{fig:structure}) with a depth of six, such that the output state perfectly matches the tree layout, in order to showcase the potential of our simulator.

We compared the running times of the MPS and TTN simulators to validate our idea.
Correctness is confirmed by contracting the networks to the full statevector and comparing them with a traditional statevector simulator.

One takeaway from this initial trial is the impact of the relation between the number of clusters and the number of leaves in a cluster.
Too many clusters introduce overhead in intermediate nodes which have to be normalized each time a gate is applied.
Only a few but bigger clusters require too much calculation on the individual normalization.
For different purposes, these two aspects have to be fine-tuned to get the best result.
Also, compared to applying gates, the structure search has no real impact on the performance.

\subsection{Circuits}

\begin{figure}
    \centering
    \begin{tikzpicture}[>=stealth, scale=0.9]
        \begin{scope}[shift={(3,4.5)}]
            \node[circle, draw, thick, minimum size=8, inner sep=0, fill=\nodecolor] (A) at (0,0) {};
            \node[circle, draw, thick, minimum size=8, inner sep=0, fill=\nodecolor] (B) at (0.5,0) {};
            \node[circle, draw, thick, minimum size=8, inner sep=0, fill=\nodecolor] (C) at (1,0) {};
            \node[circle, draw, thick, minimum size=8, inner sep=0, fill=\nodecolor] (D) at (1.5,0) {};
            \node[circle, draw, thick, minimum size=8, inner sep=0, fill=\nodecolor] (E) at (0,-0.5) {};
            \node[circle, draw, thick, minimum size=8, inner sep=0, fill=\nodecolor] (F) at (0.5,-0.5) {};
            \node[circle, draw, thick, minimum size=8, inner sep=0, fill=\nodecolor] (G) at (1,-0.5) {};
            \node[circle, draw, thick, minimum size=8, inner sep=0, fill=\nodecolor] (H) at (1.5,-0.5) {};
            \node[circle, draw, thick, minimum size=8, inner sep=0, fill=\nodecolor] (I) at (0,-1) {};
            \node[circle, draw, thick, minimum size=8, inner sep=0, fill=\nodecolor] (J) at (0.5,-1) {};
            \node[circle, draw, thick, minimum size=8, inner sep=0, fill=\nodecolor] (K) at (1,-1) {};
            \node[circle, draw, thick, minimum size=8, inner sep=0, fill=\nodecolor] (L) at (1.5,-1) {};
            \node[circle, draw, thick, minimum size=8, inner sep=0, fill=\nodecolor] (M) at (0,-1.5) {};
            \node[circle, draw, thick, minimum size=8, inner sep=0, fill=\nodecolor] (N) at (0.5,-1.5) {};
            \node[circle, draw, thick, minimum size=8, inner sep=0, fill=\nodecolor] (O) at (1,-1.5) {};
            \node[circle, draw, thick, minimum size=8, inner sep=0, fill=\nodecolor] (P) at (1.5,-1.5) {};
            \draw[thick] (A) -- (B) -- (C) -- (D);
            \draw[thick] (E) -- (F) -- (G) -- (H);
            \draw[thick] (I) -- (J) -- (K) -- (L);
            \draw[thick] (M) -- (N) -- (O) -- (P);
            \draw[thick] (A) -- (E) -- (I) -- (M);
            \draw[thick] (B) -- (F) -- (J) -- (N);
            \draw[thick] (C) -- (G) -- (K) -- (O);
            \draw[thick] (D) -- (H) -- (L) -- (P);

            \begin{scope}[on background layer]
                \filldraw[thick,fill=mgreen!20,draw=mgreen!50] (-0.25,0) arc (180:90:0.25) -- (0.5,0.25) arc (90:-90:0.25)  arc (90:180:0.25) arc (0:-180:0.25) -- cycle;
                \filldraw[thick,fill=mpurple!20,draw=mpurple!50] (1.5,0.25) arc (90:0:0.25) -- (1.75,-0.5) arc (0:-180:0.25) arc (0:90:0.25) arc (270:90:0.25) -- cycle;
                \filldraw[thick,fill=morange!20,draw=morange!50] (-0.25,-1.5) arc (180:270:0.25) -- (0.5,-1.75) arc (-90:90:0.25) arc (270:180:0.25) arc (0:180:0.25) -- cycle;
                \filldraw[thick,fill=mcyan!20,draw=mcyan!50] (1.75,-1.5) arc (0:-90:0.25) -- (1,-1.75) arc (270:90:0.25) arc (-90:0:0.25) arc (180:0:0.25)  -- cycle;
            \end{scope}
            
            \node at (0.75, -2) {$\downarrow$};
        \end{scope}
        \node[circle, draw, thick, minimum size=8, inner sep=0, fill=\nodecolor] (A) at (0,0) {};
        \node[circle, draw, thick, minimum size=8, inner sep=0, fill=\nodecolor] (B) at (0.5,0) {};
        \node[circle, draw, thick, minimum size=8, inner sep=0, fill=\nodecolor] (C) at (1,0) {};
        \node[circle, draw, thick, minimum size=8, inner sep=0, fill=\nodecolor] (D) at (1.5,0) {};

        \node[circle, draw, thick, minimum size=8, inner sep=0, fill=\nodecolor] (E) at (2,0) {};
        \node[circle, draw, thick, minimum size=8, inner sep=0, fill=\nodecolor] (F) at (2.5,0) {};
        \node[circle, draw, thick, minimum size=8, inner sep=0, fill=\nodecolor] (G) at (3,0) {};
        \node[circle, draw, thick, minimum size=8, inner sep=0, fill=\nodecolor] (H) at (3.5,0) {};

        \node[circle, draw, thick, minimum size=8, inner sep=0, fill=\nodecolor] (I) at (4,0) {};
        \node[circle, draw, thick, minimum size=8, inner sep=0, fill=\nodecolor] (J) at (4.5,0) {};
        \node[circle, draw, thick, minimum size=8, inner sep=0, fill=\nodecolor] (K) at (5,0) {};
        \node[circle, draw, thick, minimum size=8, inner sep=0, fill=\nodecolor] (L) at (5.5,0) {};
        
        \node[circle, draw, thick, minimum size=8, inner sep=0, fill=\nodecolor] (M) at (6,0) {};
        \node[circle, draw, thick, minimum size=8, inner sep=0, fill=\nodecolor] (N) at (6.5,0) {};
        \node[circle, draw, thick, minimum size=8, inner sep=0, fill=\nodecolor] (O) at (7,0) {};
        \node[circle, draw, thick, minimum size=8, inner sep=0, fill=\nodecolor] (P) at (7.5,0) {};

        \node[circle, draw, thick, minimum size=8, inner sep=0, fill=\nodecolor] (T1) at (0.5,0.5) {};
        \node[circle, draw, thick, minimum size=8, inner sep=0, fill=\nodecolor] (T2) at (3,0.5) {};
        \node[circle, draw, thick, minimum size=8, inner sep=0, fill=\nodecolor] (T3) at (4.5,0.5) {};
        \node[circle, draw, thick, minimum size=8, inner sep=0, fill=\nodecolor] (T4) at (7,0.5) {};
        \node[circle, draw, thick, minimum size=8, inner sep=0, fill=\nodecolor] (T5) at (1.25,1) {};
        \node[circle, draw, thick, minimum size=8, inner sep=0, fill=\nodecolor] (T6) at (2.25,1) {};
        \node[circle, draw, thick, minimum size=8, inner sep=0, fill=\nodecolor] (T7) at (5.25,1) {};
        \node[circle, draw, thick, minimum size=8, inner sep=0, fill=\nodecolor] (T8) at (6.25,1) {};
        \node[circle, draw, thick, minimum size=8, inner sep=0, fill=\nodecolor] (T9) at (3.75,2) {};
        \node (A1) at (0,-0.5) {};
        \node (B1) at (0.5,-0.5) {};
        \node (C1) at (1,-0.5) {};
        \node (D1) at (1.5,-0.5) {};
        \node (E1) at (2,-0.5) {};
        \node (F1) at (2.5,-0.5) {};
        \node (G1) at (3,-0.5) {};
        \node (H1) at (3.5,-0.5) {};
        \node (I1) at (4,-0.5) {};
        \node (J1) at (4.5,-0.5) {};
        \node (K1) at (5,-0.5) {};
        \node (L1) at (5.5,-0.5) {};
        \node (M1) at (6,-0.5) {};
        \node (N1) at (6.5,-0.5) {};
        \node (O1) at (7,-0.5) {};
        \node (P1) at (7.5,-0.5) {};
        \draw[thick] (A1) -- (A) -- (T1) -- (T5)  -- (T9);
        \draw[thick] (B1) -- (B) -- (T1);
        \draw[thick] (C1) -- (C) -- (T1);
        \draw[thick] (D1) -- (D) -- (T5);
        \draw[thick] (E1) -- (E) -- (T6);
        \draw[thick] (F1) -- (F) -- (T2) -- (T6) -- (T9);
        \draw[thick] (G1) -- (G) -- (T2);
        \draw[thick] (H1) -- (H) -- (T2);
        \draw[thick] (I1) -- (I) -- (T3) -- (T7) -- (T9);
        \draw[thick] (J1) -- (J) -- (T3);
        \draw[thick] (K1) -- (K) -- (T3);
        \draw[thick] (L1) -- (L) -- (T7);
        \draw[thick] (M1) -- (M) -- (T8);
        \draw[thick] (N1) -- (N) -- (T4) -- (T8) -- (T9);
        \draw[thick] (O1) -- (O) -- (T4);
        \draw[thick] (P1) -- (P) -- (T4);
        \begin{scope}[on background layer]
            \filldraw[thick,fill=mgreen!20,draw=mgreen!50] (0,-0.25) arc (270:90:0.25) -- (1,0.25) arc (90:-90:0.25) -- cycle;
            \filldraw[thick,fill=mpurple!20,draw=mpurple!50] (2.5,-0.25) arc (270:90:0.25) -- (3.5,0.25) arc (90:-90:0.25) -- cycle;
            \filldraw[thick,fill=morange!20,draw=morange!50] (4,-0.25) arc (270:90:0.25) -- (5,0.25) arc (90:-90:0.25) -- cycle;
            \filldraw[thick,fill=mcyan!20,draw=mcyan!50] (6.5,-0.25) arc (270:90:0.25) -- (7.5,0.25) arc (90:-90:0.25) -- cycle;
        \end{scope}
    \end{tikzpicture}
    \caption{$4\times4$-qubit lattice with a nearest neighbor gate pattern, and corresponding tree architecture for representing the output quantum state.}
    \label{fig:sycamore}
\end{figure}
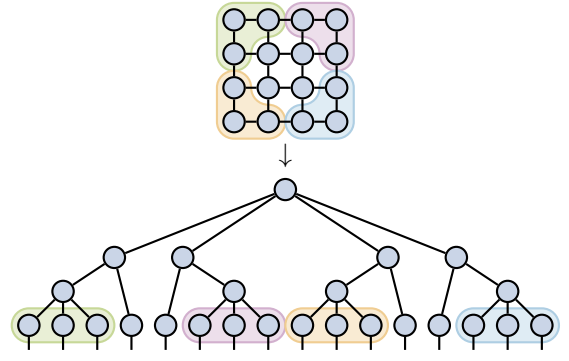

For our purposes, quantum circuits can be divided into two different categories, based on how well they can be clustered into a tree layout.
Some circuits are not suitable due to their connectivity pattern, for example, the quantum Fourier transform with all-to-all connectivity or circuits with tight nearest-neighbor connectivity.
This category is represented by the \emph{lattice} circuit in the simulations.
Fig.~\ref{fig:sycamore} shows the resulting tree for a $4\times4$ qubit lattice and gate pattern based on the Google Sycamore experiment \cite{Arute2019} for randomized circuits.
They specify a pattern of nearest neighbor gate activations alternating on each row and column, for an exact specification refer to \cite{Arute2019}.
All gates are chosen randomly from a set of predefined gates.

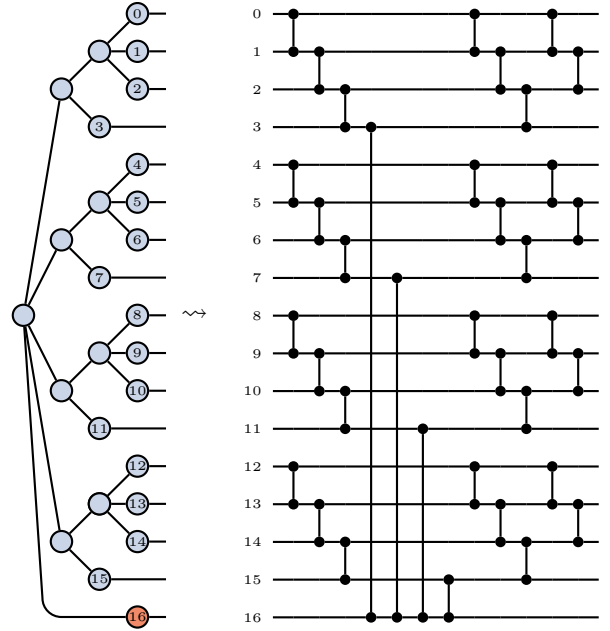
\begin{figure}
    \begin{tikzpicture}[xscale=-1,rotate=-90]
        \node[circle, draw, thick, minimum size=8, inner sep=0, fill=\nodecolor] (A) at (0,0) {\tiny 0};
        \node[circle, draw, thick, minimum size=8, inner sep=0, fill=\nodecolor] (B) at (0.5,0) {\tiny 1};
        \node[circle, draw, thick, minimum size=8, inner sep=0, fill=\nodecolor] (C) at (1,0) {\tiny 2};
        \node[circle, draw, thick, minimum size=8, inner sep=0, fill=\nodecolor] (E) at (1.5,0.5) {\tiny 3};
        
        \node[circle, draw, thick, minimum size=8, inner sep=0, fill=\nodecolor] (F) at (2,0) {\tiny 4};
        \node[circle, draw, thick, minimum size=8, inner sep=0, fill=\nodecolor] (G) at (2.5,0) {\tiny 5};
        \node[circle, draw, thick, minimum size=8, inner sep=0, fill=\nodecolor] (I) at (3,0) {\tiny 6};
        \node[circle, draw, thick, minimum size=8, inner sep=0, fill=\nodecolor] (J) at (3.5,0.5) {\tiny 7};

        \node[circle, draw, thick, minimum size=8, inner sep=0, fill=\nodecolor] (K) at (4,0) {\tiny 8};
        \node[circle, draw, thick, minimum size=8, inner sep=0, fill=\nodecolor] (M) at (4.5,0) {\tiny 9};
        \node[circle, draw, thick, minimum size=8, inner sep=0, fill=\nodecolor] (N) at (5,0) {\tiny 10};
        \node[circle, draw, thick, minimum size=8, inner sep=0, fill=\nodecolor] (O) at (5.5,0.5) {\tiny 11};

        \node[circle, draw, thick, minimum size=8, inner sep=0, fill=\nodecolor] (D) at (6,0) {\tiny 12};
        \node[circle, draw, thick, minimum size=8, inner sep=0, fill=\nodecolor] (H) at (6.5,0) {\tiny 13};
        \node[circle, draw, thick, minimum size=8, inner sep=0, fill=\nodecolor] (L) at (7,0) {\tiny 14};
        \node[circle, draw, thick, minimum size=8, inner sep=0, fill=\nodecolor] (P) at (7.5,0.5) {\tiny 15};

        \node[circle, draw, thick, minimum size=8, inner sep=0, fill=\gatecolor] (Q) at (8,0) {\tiny 16};

        \node[circle, draw, thick, minimum size=8, inner sep=0, fill=\nodecolor] (T1) at (0.5,0.5) {};
        \node[circle, draw, thick, minimum size=8, inner sep=0, fill=\nodecolor] (T2) at (2.5,0.5) {};
        \node[circle, draw, thick, minimum size=8, inner sep=0, fill=\nodecolor] (T3) at (4.5,0.5) {};
        \node[circle, draw, thick, minimum size=8, inner sep=0, fill=\nodecolor] (T4) at (6.5,0.5) {};
        \node[circle, draw, thick, minimum size=8, inner sep=0, fill=\nodecolor] (T5) at (6.5,0.5) {};
        \node[circle, draw, thick, minimum size=8, inner sep=0, fill=\nodecolor] (T6) at (4,1.5) {};
        
        \node[circle, draw, thick, minimum size=8, inner sep=0, fill=\nodecolor] (T7) at (1,1) {};
        \node[circle, draw, thick, minimum size=8, inner sep=0, fill=\nodecolor] (T8) at (3,1) {};
        \node[circle, draw, thick, minimum size=8, inner sep=0, fill=\nodecolor] (T9) at (5,1) {};
        \node[circle, draw, thick, minimum size=8, inner sep=0, fill=\nodecolor] (T10) at (7,1) {};

        \node at (4,-0.75) {$\rightsquigarrow$};
        \node (A1) at (0,-0.5) {};
        \node (B1) at (0.5,-0.5) {};
        \node (C1) at (1,-0.5) {};
        \node (D1) at (6,-0.5) {};
        \node (E1) at (1.5,-0.5) {};
        \node (F1) at (2,-0.5) {};
        \node (G1) at (2.5,-0.5) {};
        \node (H1) at (6.5,-0.5) {};
        \node (I1) at (3,-0.5) {};
        \node (J1) at (3.5,-0.5) {};
        \node (K1) at (4,-0.5) {};
        \node (L1) at (7,-0.5) {};
        \node (M1) at (4.5,-0.5) {};
        \node (N1) at (5,-0.5) {};
        \node (O1) at (5.5,-0.5) {};
        \node (P1) at (7.5,-0.5) {};
        \node (Q1) at (8,-0.5) {};
        \draw[thick] (A1) -- (A) -- (T1) -- (T7) -- (T6);
        \draw[thick] (B1) -- (B) -- (T1);
        \draw[thick] (C1) -- (C) -- (T1);
        \draw[thick] (D1) -- (D) -- (T4) -- (T10) -- (T6);
        \draw[thick] (E1) -- (E) -- (T7);
        \draw[thick] (F1) -- (F) -- (T2) -- (T8) -- (T6);
        \draw[thick] (G1) -- (G) -- (T2);
        \draw[thick] (H1) -- (H) -- (T4);
        \draw[thick] (I1) -- (I) -- (T2);
        \draw[thick] (J1) -- (J) -- (T8);
        \draw[thick] (K1) -- (K) -- (T3) -- (T9) -- (T6);
        \draw[thick] (L1) -- (L) -- (T4);
        \draw[thick] (M1) -- (M) -- (T3);
        \draw[thick] (N1) -- (N) -- (T3);
        \draw[thick] (O1) -- (O) -- (T9);
        \draw[thick] (P1) -- (P) -- (T10);
        \draw[thick] (Q1) -- (Q) -- (8,1) arc (0:90:0.25) -- (T6);
    \end{tikzpicture}
    \raisebox{4.15cm}{%
    \begin{quantikz}[row sep={0.5cm,between origins}, column sep=0.2cm]
        \lstick{\tiny 0}  & \ctrl{1}      & \qw           & \qw           & \qw           & \qw           & \qw           & \qw         & \ctrl{1}      & \qw           & \qw        & \ctrl{1}      & \qw           & \qw  \\
        \lstick{\tiny 1}  & \control{}    & \ctrl{1}      & \qw           & \qw           & \qw           & \qw           & \qw         & \control{}    & \ctrl{1}      & \qw        & \control{}    & \ctrl{1}      & \qw  \\
        \lstick{\tiny 2}  & \qw           & \control{}    & \ctrl{1}      & \qw           & \qw           & \qw           & \qw         & \qw           & \control{}    & \ctrl{1}   & \qw           & \control{}    & \qw  \\
        \lstick{\tiny 3}  & \qw           & \qw           & \control{}    & \ctrl{13}     & \qw           & \qw           & \qw         & \qw           & \qw           & \control{} & \qw           & \qw           & \qw  \\
        \lstick{\tiny 4}  & \ctrl{1}      & \qw           & \qw           & \qw           & \qw           & \qw           & \qw         & \ctrl{1}      & \qw           & \qw        & \ctrl{1}      & \qw           & \qw  \\
        \lstick{\tiny 5}  & \control{}    & \ctrl{1}      & \qw           & \qw           & \qw           & \qw           & \qw         & \control{}    & \ctrl{1}      & \qw        & \control{}    & \ctrl{1}      & \qw  \\
        \lstick{\tiny 6}  & \qw           & \control{}    & \ctrl{1}      & \qw           & \qw           & \qw           & \qw         & \qw           & \control{}    & \ctrl{1}   & \qw           & \control{}    & \qw  \\
        \lstick{\tiny 7}  & \qw           & \qw           & \control{}    & \qw           & \ctrl{9}      & \qw           & \qw         & \qw           & \qw           & \control{} & \qw           & \qw           & \qw  \\
        \lstick{\tiny 8}  & \ctrl{1}      & \qw           & \qw           & \qw           & \qw           & \qw           & \qw         & \ctrl{1}      & \qw           & \qw        & \ctrl{1}      & \qw           & \qw  \\
        \lstick{\tiny 9}  & \control{}    & \ctrl{1}      & \qw           & \qw           & \qw           & \qw           & \qw         & \control{}    & \ctrl{1}      & \qw        & \control{}    & \ctrl{1}      & \qw  \\
        \lstick{\tiny 10} & \qw           & \control{}    & \ctrl{1}      & \qw           & \qw           & \qw           & \qw         & \qw           & \control{}    & \ctrl{1}   & \qw           & \control{}    & \qw  \\
        \lstick{\tiny 11} & \qw           & \qw           & \control{}    & \qw           & \qw           & \ctrl{5}      & \qw         & \qw           & \qw           & \control{} & \qw           & \qw           & \qw  \\
        \lstick{\tiny 12} & \ctrl{1}      & \qw           & \qw           & \qw           & \qw           & \qw           & \qw         & \ctrl{1}      & \qw           & \qw        & \ctrl{1}      & \qw           & \qw  \\
        \lstick{\tiny 13} & \control{}    & \ctrl{1}      & \qw           & \qw           & \qw           & \qw           & \qw         & \control{}    & \ctrl{1}      & \qw        & \control{}    & \ctrl{1}      & \qw  \\
        \lstick{\tiny 14} & \qw           & \control{}    & \ctrl{1}      & \qw           & \qw           & \qw           & \qw         & \qw           & \control{}    & \ctrl{1}   & \qw           & \control{}    & \qw  \\
        \lstick{\tiny 15} & \qw           & \qw           & \control{}    & \qw           & \qw           & \qw           & \ctrl{1}    & \qw           & \qw           & \control{} & \qw           & \qw           & \qw  \\
        \lstick{\tiny 16} & \qw           & \qw           & \qw           & \control{}    & \control{}    & \control{}    & \control{}  & \qw           & \qw           & \qw        & \qw           & \qw           & \qw  
    \end{quantikz}}
\caption{Well-structured circuit pattern resulting from a clusterable tree. The highlighted subtree contains all connections crossing boundaries.}
\label{fig:structure}
\end{figure}

In some circuits, the output state almost perfectly matches the tree layout.
The \emph{tree-like} circuit showcases the potential of our simulator without depending on the recursive nature.
A pattern is created from a highly entangled cluster with four qubits, which is repeated any number of times, similar to the example in Fig.~\ref{fig:example_connectivity}.
Gates crossing cluster boundaries are included, but kept under the $D_{\max}$ threshold by setting $k_G = 2$, only connecting to one central location. 
Fig.~\ref{fig:structure} shows an instance with four clusters and 17 qubits, derived from the given tree.

\subsection{Experiments and results}

In this section, we discuss the experiments and results we achieve by running them.
We also provide a rationale for why approximation techniques (based on truncating singular values) are not useful in this regime.

\subsubsection{Wall-clock timing}

\begin{figure}
    \includegraphics[width=\columnwidth]{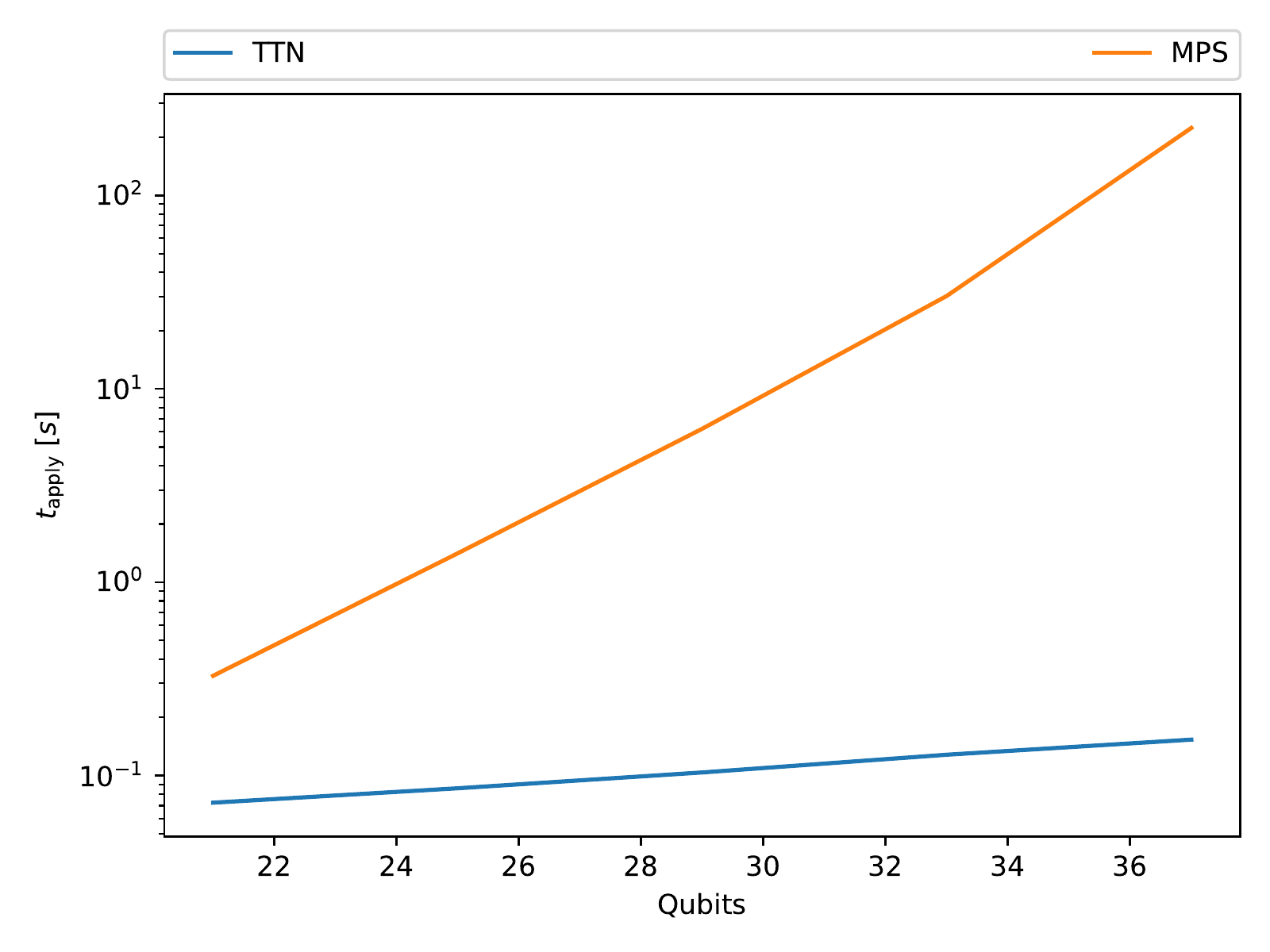}
    \caption{Wall-clock time of applying gates and re-orthonormalization, comparing a MPS with a TTN representation of the quantum state for the \emph{tree-like} circuit.}
    \label{fig:experiment3}
\end{figure}

The experiment is conducted for the \emph{tree-like} circuit and scaled up to a maximum of 37 qubits which just fits into 256 GB of swap space.
We simply measure the wall-clock time of simulating the complete circuit.
The results from this experiment, shown in Fig.~\ref{fig:experiment3}, indicate that the TTN outperforms the MPS.
The memory requirements described in Sec.~\ref{sec:memory_cost} can be fulfilled for certain tree structures.
For some cluster sizes, the condition from Eq.~\eqref{eq:edge_gates_restriction} holds, which results in a decrease in internal bond dimensions.
Timings are highly dependent on the real tensor structure.
Our algorithm is tailored to provide a generally useful structure but this can be optimized.
The implementation provided in Sec.~\ref{sec:experiments} allows overwriting the structure by the user if necessary 

\begin{figure}
    \includegraphics[width=\columnwidth]{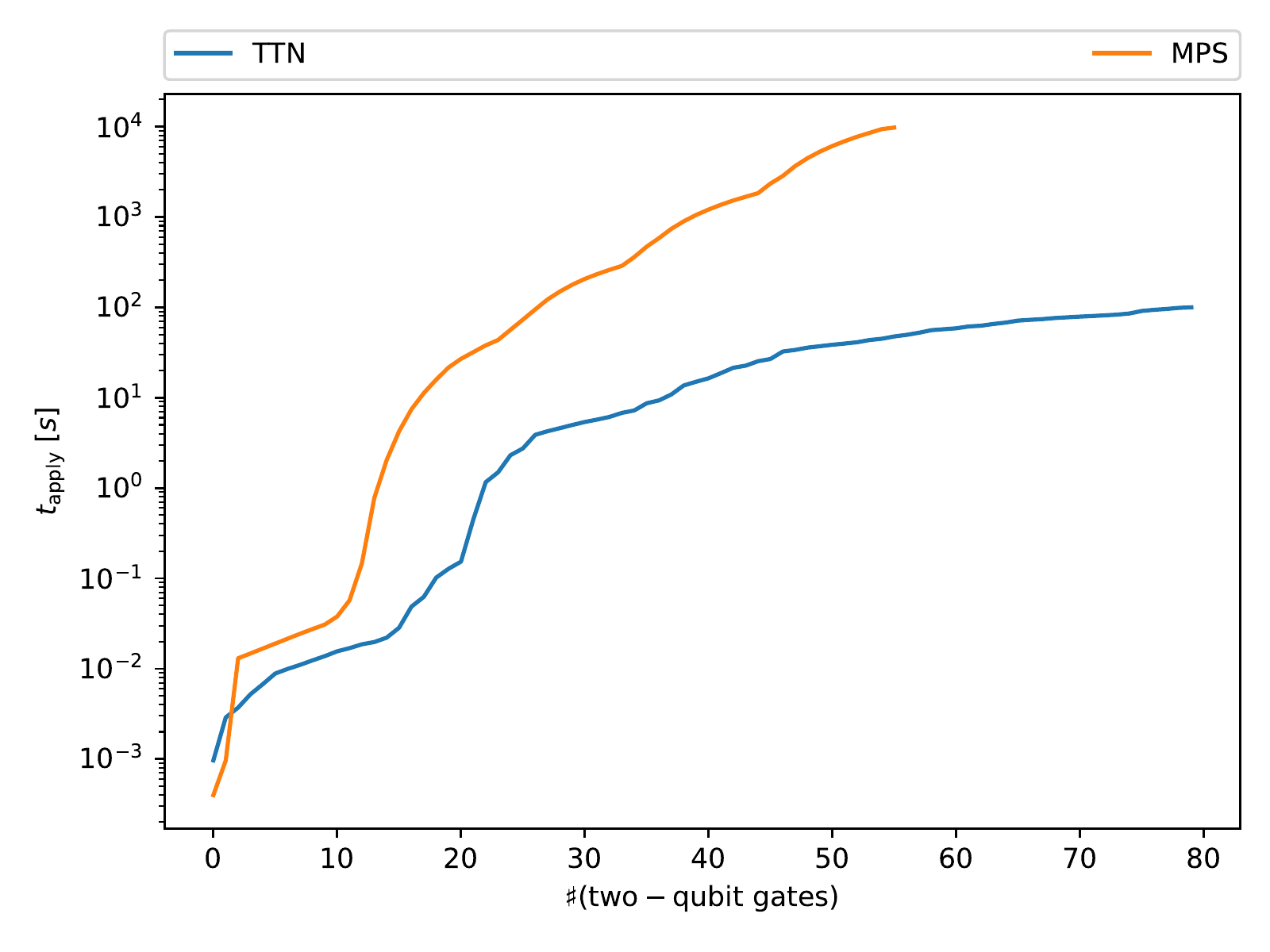}
    \caption{Accumulated wall-clock time of applying gates and re-orthonormalization, on the \emph{lattice} circuit (for $25$ qubits). For gate 57 the MPS fails to normalize.}
    \label{fig:sycamore_experiment3}
\end{figure}

One unexpected outcome is the performance of TTNs on the \emph{lattice} circuits.
In a further trial on a $5\times5$ \emph{lattice} the MPS fails to normalize some nodes. 
The Fortran BLAS implementation which we use for SVD and QR-decomposition does not converge due to the high internal dimensions.
Fig.~\ref{fig:sycamore_experiment3} shows the results up to the point of failure.
In comparison, the TTN delivers a result in a reasonable time.
A possible explanation for this could be the shorter path distance between some leaves, which means a fewer number of normalizations.

\subsubsection{Scaling for a large number of qubits}

To circumvent numerical issues, we also \emph{simulate} the gate application and orthonormalization procedure in an additional experiment.
These ``dry-runs'' allow for calculating the internal bond dimensions without performing the actual tensor operations.
This enables reasoning for qubit numbers above the classically possible threshold.  
For the comparison, we measure two aspects.
Firstly, the time of the circuit calculation which includes finding the tree structure and applying the gates with a renormalization sweep afterward.
Secondly, two metrics for bond dimensions are considered: the maximum internal bond dimension $D_{\max} = \max_{T \in \text{TN}} \max_i(\dim_i(T))$, and as a measure of the memory requirements the overall number of entries, i.e., the sum over all internal tensor sizes: $M = \sum_{T \in \text{TN}} \prod_i \dim_i(T)$. 
In this notation, TN refers to either the MPS or the TTN.
We also used different initialization procedures for individual clusters not mentioned here.
These variations represent restrictions imposed on the trees.

\begin{figure}
    \subfloat[\emph{tree-like} circuit]{\includegraphics[width=\columnwidth]{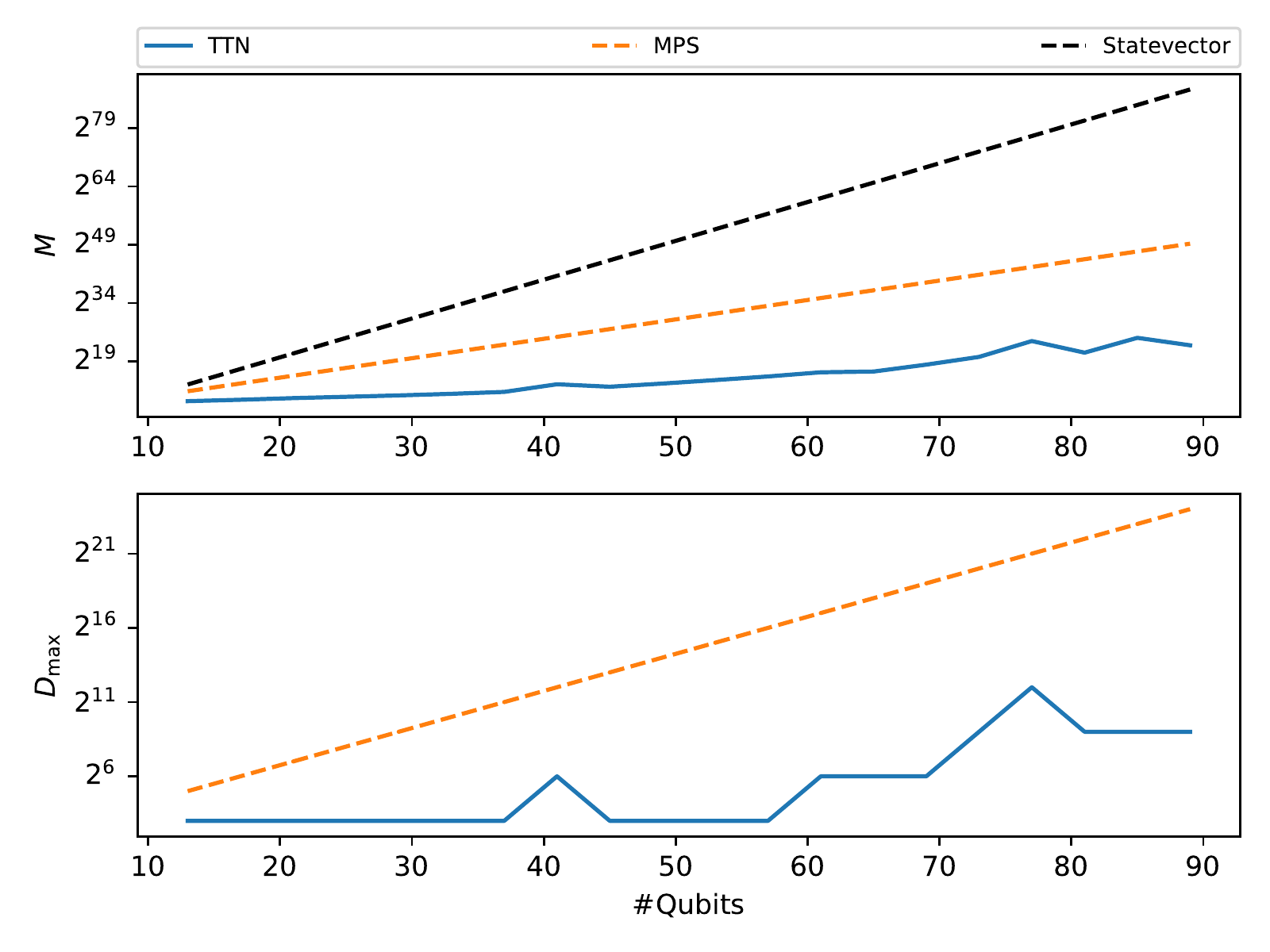}} \\
    \subfloat[\emph{lattice} circuit]  {\includegraphics[width=\columnwidth]{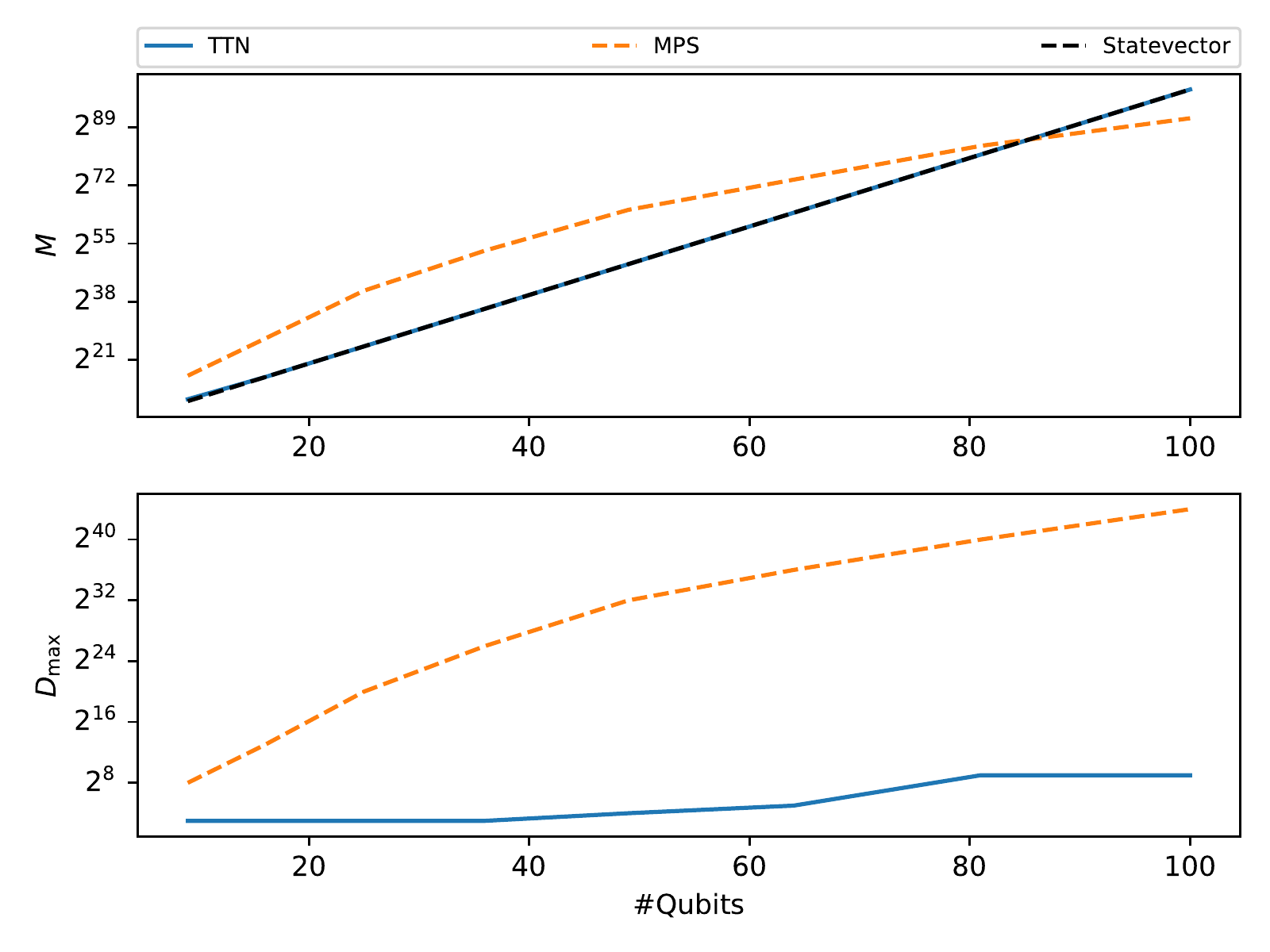}}
    \caption{Results of the \emph{dry-runs} with different settings for cluster generation compared to the MPS. 
    The non-monotonicity for the \emph{tree-like} circuit is caused by the heuristics in the tree creation. 
    In some cases, the algorithm fails to find a good clustering, namely for \emph{41} and \emph{77} qubits.}
    \label{fig:experiment2}
\end{figure}

The dry-run experiment provides insight into the scalability of the TTN approach.
On circuits with a large number of qubits, MPS and TTN scale similarly in regards to $M$.
Fig.~\ref{fig:experiment2} shows the results for sizes up to 100 qubits.
By construction, the circuit is shallow to conform with Eq.~\eqref{eq:edge_gates_restriction}.
Most advantages can only be achieved if the cluster boundaries are not crossed which will inevitably happen for deep circuits.
In the \emph{lattice} circuit, the number of connections already exceeds the threshold and the performance deteriorates.

Another aspect that arises from the results is the versatility of the approach.
The creation and structure of the clusters can be tuned to fit different metrics.
This can also be extended to include hardware specifications if necessary.
Bad clustering, can harm the performance in certain cases.
If leaves are distributed across clusters too unevenly, the biggest cluster dominates the performance cost, also evident in Fig.~\ref{fig:experiment2}.

\subsubsection{Effect of approximations}

Similar to \cite{Zhou2020}, we examine the effect of truncating singular values on error and memory savings at levels $> \ell_{\text{cluster}}$.
In these cases, we discard the singular values $\sigma_{\text{truncated}}$, given $\sigma_1 \ge \dots \ge \sigma_{\text{threshold}} > \sigma_{\text{truncated}} \ge 0$ after normalization.
The fidelity of the state after truncation is quantified by the overlap error, which is equivalent to the trace distance when considering pure states and is far less memory intensive.
\begin{equation}
    \mathcal{E}_{\text{overlap}} = 1 - \lvert\braket{\psi_{\text{tree}} \vert \psi_{\text{exact}}} \rvert^2
\end{equation}
\begin{figure}
    \subfloat[Randomized $16$-qubit circuits]{\includegraphics[width=\columnwidth]{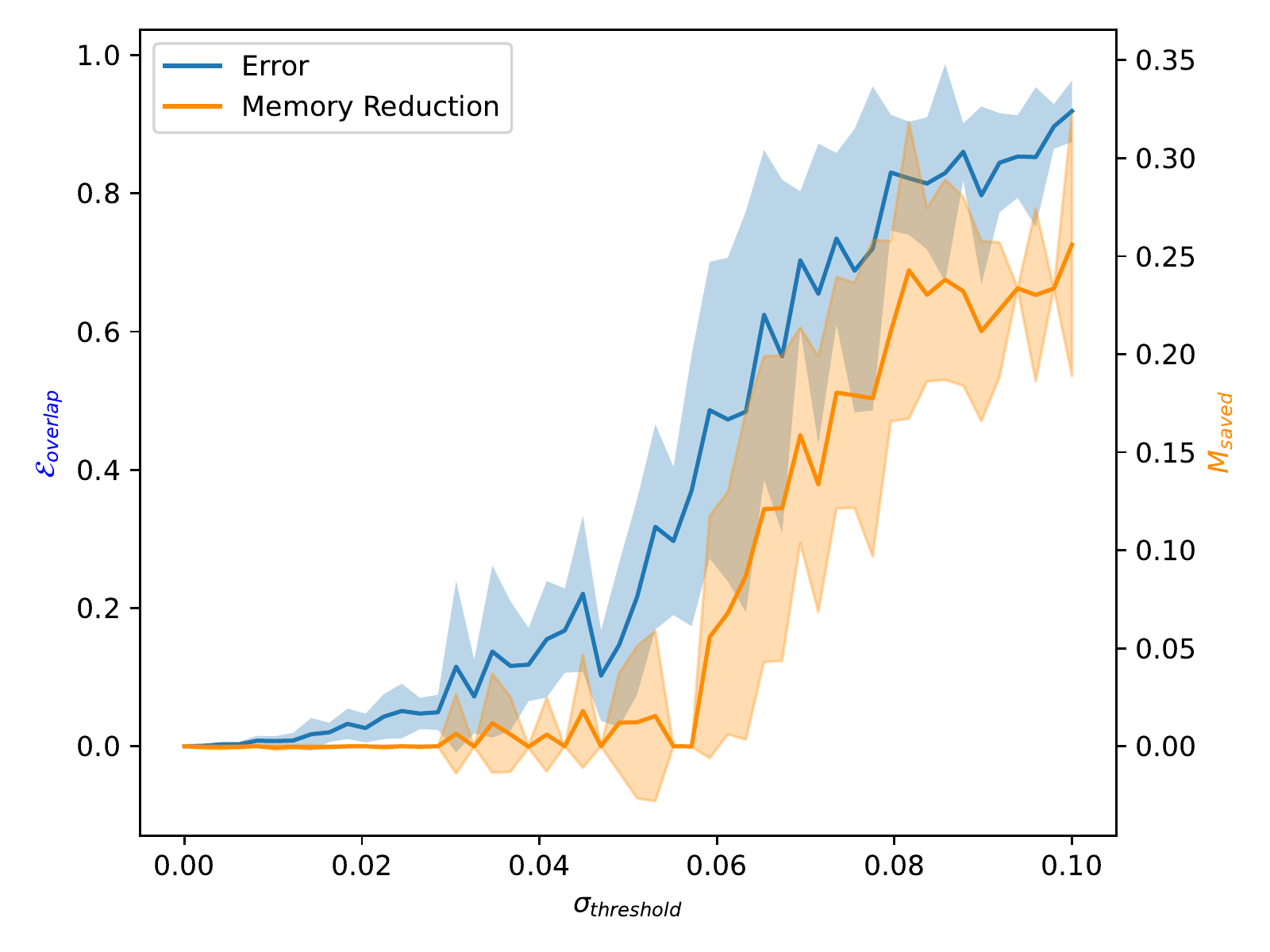}} \\
    \subfloat[Example for singular values \label{fig:svds}]{\includegraphics[width=\columnwidth]{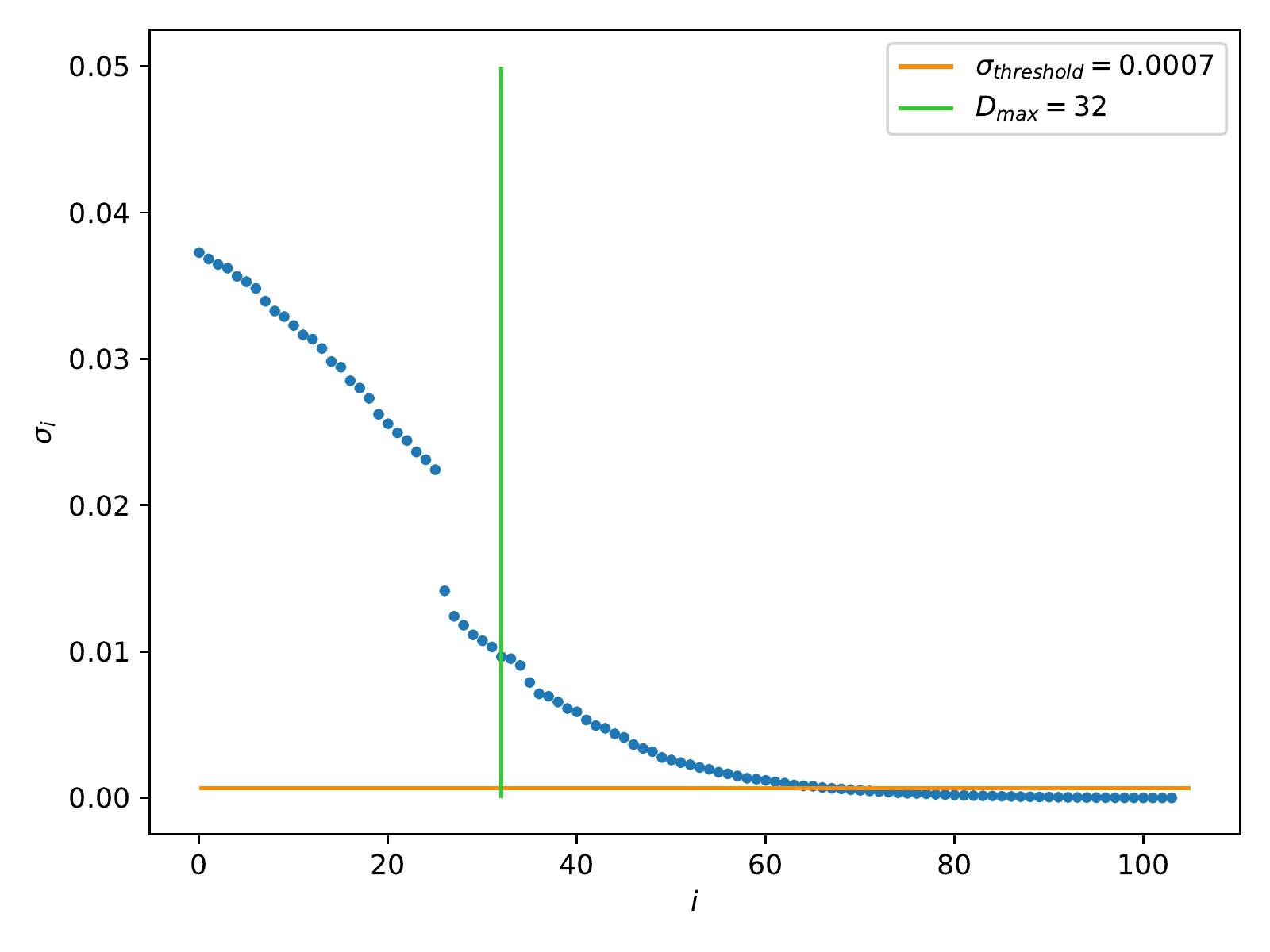}}
    \caption{Errors and the resulting saved dimensionality for our proposed truncation scheme on randomized circuits. The relative saved memory is $\propto M_{\text{saved}}= 1 -\frac{M_{\text{truncated}}}{M_{\text{perfect}}}$.
        b) gives an example for an accumulated cut of off $0.1$ which already induces an error of approximately 25\%}
    \label{fig:experiment4}
\end{figure}

We use the \emph{lattice}-circuit with randomized nearest-neighbor interactions and run $10$ trials for each tolerance.
For small values of $\sigma_{\text{threshold}}$, the error is reasonable, but it barely results in any saved dimensionality.
We see a significant increase in error given a singular value threshold of $~7e-4$, which also corresponds to an accumulated cutoff of $0.1$, see Fig.~\ref{fig:svds}.
The resulting saved memory is not worth the tradeoff.
Meaningful improvements can only be exchanged with a larger error rate.
This is a side effect of the proposed cluster construction.
In our truncation scheme, this can be explained by two factors:
\begin{enumerate}
    \item The relative size of the singular values.
    \item The number of necessary truncations.
\end{enumerate}
The first aspect is mainly caused by the chosen cluster size.
Smaller clusters perform better, but on smaller tensors truncating even a small number of singular values introduces significant errors.
The second aspect arises from the normalization procedure which requires renormalization after each gate is applied.
Once a given $D_{\max}$ is reached, each additional gate -would need truncation.
Since we cannot avoid either of the two scenarios, we found the resulting error to be unacceptably large in this experiment.
The only exceptions are circuits where cluster boundaries are no longer crossed after a fixed point in time.

Similarly, problems arise when controlling the bond dimension directly.
Fixing $D_{\max}$ to a smaller bound, suffers the same pitfalls as before.
Fig.~\ref{fig:svds} showcases the problem as one truncation can already produce a high error.
Analogously, each following gate potentially introduces a similar error.
It is open to see if other combinations of construction and truncation schemes might be more successful.
For now, other approximation techniques remain superior in terms of the error-to-performance tradeoff.

\section{Conclusions and outlook}

In this work, we introduce a novel method to perform circuit simulation that exploits circuit structure using a TTN representation, which can be handled efficiently on a classical computer. A method to generate an advantageous tree structure for the initial quantum state which restricts bond dimension growth during circuit simulation is presented. Numerical simulations were run, demonstrating an advantage in overall simulation time (wall-clock time) between a naive MPS formulation and the presented method for specific circuits. Additionally, a specific circuit layout is presented in which MPSs would see exponential bond dimension growth, while states in the TTN representation would retain a uniformly bounded bond dimension.

A practical use case for the introduced method could be the now-ubiquitous Quantum Approximate Optimization Algorithm (QAOA), which is used to solve combinatorial optimization problems \cite{Farhi2014}.
In this algorithm, the optimal solution to the problem is formulated as the ground state of a constructed Hamiltonian.
The time evolution of such Hamiltonians often requires irregular connectivity, for which our TTN approach is suitable, especially as the layers of irregularly connected gates are repeated several times.
One classical example is the Max-Cut problem, which is described in detail in~\cite{Farhi2014}, which, when mapped to a graph, exhibits the same connectivity as the original tree structure.

Another instance where TTNs might be useful is in entangled ancilla protocols as in \cite{Schiffer2021}.
Depending on the construction, multiple ancillae are entangled with a low number of gates and then only connected to their respective circuit.
The problem remains intractable on classical computers for large circuit sizes, but the simulator could be used for confirming correctness on smaller problem instances. 

TTNs have also shown to be a valuable technique for many-body physics \cite{Gerster2014}.
Most approaches and algorithms studied are restricted to binary TTNs.
However, we are not aware of any approaches that allow for an arbitrary \emph{m-ary} or non-regular trees and exploration in a more generalized tree structure could yield interesting results.
In quantum chemistry \cite{Murg2015, Szalay2015} such TTNs have already been used and the importance of topology optimization is shown \cite{Szalay2015}.
Our algorithm is designed to work with arbitrary similarity functions but is restricted to quantum circuits.
A more general approach could incorporate entanglement or utilize problem-specific similarity functions.
For example, \cite{Gerster2014} considers a fixed bond dimension and provides an algorithm to find an advantageous binary TTN.
One could lift this restriction to binary TTNs in certain well-structured regions to allow for a more compact representation.
In the extreme, optimization for each tensor might be possible which could allow generalizing to more complicated topologies. 

From an algorithmic perspective, several avenues of future work still exist: the current algorithm assumes that two-qubit gates always introduce the maximum amount of entanglement, resulting in the restriction Eq.~\eqref{eq:edge_gates_restriction}.
The introduced method would benefit from a ``look-ahead'' estimation of entanglement generated by two-qubit gates; which is tractable if we reduce simulated circuits to more ubiquitous gate sets.
Another promising generalization would be dynamically adapting the tree layout, i.e., creating or merging subtrees and branches when applying gates.

While not further explored in this work, an additional advantage is the possibility of interpretable and quantifiable compression by truncating the singular values of internal bonds.
Namely, the singular values characterize the entanglement between subsystems and thus have a physical interpretation.
On the other hand, the effects of truncating singular values within a general tensor network contraction are typically less predictable, and hence such contractions are usually avoided.

We hope that our algorithm for the tree structure search and theoretical analysis is applicable and insightful for other use cases of TTNs as well, e.g., condensed matter or quantum chemistry simulations.

\acknowledgments

This project is supported by the Federal Ministry for Economic Affairs and Climate Action on the basis of a decision by the German Bundestag through project QuaST, as well as by the Bavarian Ministry of Economic Affairs, Regional Development and Energy with funds from the Hightech Agenda Bayern.

We also thank the Munich Center for Quantum Science and Technology (MCQST) and the Munich Quantum Valley (MQV) initiative.

\bibliographystyle{quantum}
\bibliography{bibliography}

\end{document}